\documentclass[a4paper,11pt]{article}
\pdfoutput=1 

\usepackage{jcappub} 

\usepackage[T1]{fontenc} 

\newcommand{\bq}{\boldsymbol q}
\newcommand{\mo}{\mathcal{O}}

\newcommand{\bx}{\boldsymbol x}
\newcommand{\bv}{\boldsymbol v}
\newcommand{\bk}{\textbf{k}}

\newcommand{\ihmpc}{\,h{\rm Mpc}^{-1}}
\newcommand{\hmsun}{\,h^{-1}M_{\odot}}

\title{\boldmath Precision Redshift-Space Galaxy Power Spectra using Zel'dovich Control Variates}

\author[a]{Joseph DeRose}
\author[b]{Shi-Fan Chen}
\author[c,d]{Nickolas Kokron}
\author[a,e,f]{Martin White}

\affiliation[a]{Physics Division, Lawrence Berkeley National Laboratory, Berkeley, CA, USA}
\affiliation[b]{Institute for Advanced Study, 1 Einstein Drive, Princeton, NJ 08540, USA}
\affiliation[c]{Department of Physics, Stanford University, 382 Via Pueblo Mall, Stanford, CA 94305, USA}
\affiliation[d]{Kavli Institute for Particle Astrophysics and Cosmology, SLAC National Accelerator Laboratory, 2575 Sand Hill Road, Menlo Park, CA 94025, USA}
\affiliation[e]{Berkeley Center for Cosmological Physics, Department of Physics, UC Berkeley, CA 94720, USA}
\affiliation[f]{Department of Physics, University of California,Berkeley, CA 94720}
\emailAdd{jderose@lbl.gov}
\emailAdd{sfschen@ias.edu}
\emailAdd{kokron@stanford.edu}
\emailAdd{mwhite@berkeley.edu}

\abstract{Numerical simulations in cosmology require trade-offs between volume, resolution and run-time that limit the volume of the Universe that can be simulated, leading to sample variance in predictions of ensemble-average quantities such as the power spectrum or correlation function(s).  Sample variance is particularly acute at large scales, which is also where analytic techniques can be highly reliable.  This provides an opportunity to combine analytic and numerical techniques in a principled way to improve the dynamic range and reliability of predictions for clustering statistics.
In this paper we extend the technique of Zel'dovich control variates, previously demonstrated for 2-point functions in real space, to reduce the sample variance in measurements of 2-point statistics of biased tracers in redshift space. We demonstrate that with this technique, we can reduce the sample variance of these statistics down to their shot-noise limit out to $k \sim 0.2\, h\rm Mpc^{-1}$. This allows a better matching with perturbative models and improved predictions for the clustering of e.g.~quasars, galaxies and neutral Hydrogen measured in spectroscopic redshift surveys at very modest computational expense.  We discuss the implementation of ZCV, give some examples and provide forecasts for the efficacy of the method under various conditions.}

\begin{document}
\maketitle
\flushbottom

\section{Introduction}
\label{sec:intro}

One of the main tools in the employ of the modern cosmologist is the numerical simulation of structure formation. Such simulations are used in a broad range of cosmological studies, notably as tests of analytical models of large-scale structure and as forward models for cosmological observations. In both cases, one wishes to reduce statistical and systematic errors on measurements made from these simulations.

In the case of virtually all numerical simulations, run-time scales proportionally to the number of resolution elements employed, raised to some power (e.g.\ \cite{Springel2021,Garrison2021,Springel2010}). Thus, for a fixed number of resolution elements, statistical and systematic errors become anti-correlated, since statistical errors scale with volume, and systematic errors scale roughly with the size of the resolution elements employed. Because of this scaling, methods for reducing simulation statistical errors without increasing simulation volume are highly desirable, as then one can decrease simulation statistical error without increasing run-time or systematic error. Furthermore, one often wishes to run simulations at a number of different cosmologies in order to perform inference tasks. In this case, reducing the statistical error of individual simulations allows for a broader range of cosmologies to be simulated at fixed computational cost, thus making the models built from said simulations more accurate.

In the context of large-scale structure studies, a number of methods for reducing the statistical error of $N$-body simulations have been developed. Perhaps most common is the `fixed amplitude' method \cite{PD96,Angulo:2016hjd}, where instead of initializing a simulation with a linear density field whose amplitudes are drawn from a Rayleigh distribution, one chooses the amplitude of each Fourier mode to be the expectation value determined from the linear power spectrum. In doing so, one can reduce the variance of statistics measured from fixed amplitude simulations, with the largest reduction in variance coming in the linear regime and degrading significantly in non-linear or stochastic regimes. In addition to this technique, one often runs a second, `paired', simulation with the same fixed amplitude initial conditions, but with modes $180^{\circ}$ out of phase with the first simulation \citep{Pontzen_2016,Angulo:2016hjd}. Doing so cancels out the next-leading-order contribution to the variance of two--point matter statistics. There is now a large literature investigating the performance of these techniques in the context of various cosmological models \cite{Avila:2022pzw}, and for various statistics beyond the matter power spectrum \cite{Pontzen_2016,Villaescusa-Navarro:2018bpd,Chuang:2018ega,Anderson:2018zkm,Klypin_2020,Zhang:2020vru,Maion22}. These techniques clearly produce non-Gaussian initial conditions, but biases incurred to two- and three-point functions from this violation have been shown to be small for many cases of interest \cite{Villaescusa-Navarro:2018bpd,Chuang:2018ega,Maion22}. Furthermore, these techniques render the simulations unusable for the estimation of covariance matrices.

More recently, the method of control variates has been introduced in the context of cosmology. Control variates have been studied for many years in the statistics literature as a method for reducing the variance of Monte Carlo estimators in general \citep{mcbook}. When estimating the mean of a random variable of interest via Monte Carlo, if there exists a correlated random variable, i.e.\ the control variate, whose mean is precisely known, then one can construct an estimator for the mean of the variable of interest that has drastically improved convergence properties. This method was introduced in the cosmology literature under the name `Convergence Acceleration by Regression and Pooling' (CARPool) in order to reduce the variance of summary statistics, such as $N$-point statistics \cite{chartier2020,Ding:2022ydj}, and covariance matrices \cite{chartier2021,Chartier:2022kjz} measured from cosmological simulations. This method does not require the simulator to alter any properties of the initial conditions, e.g. by introducing non-Gaussianity, and with a well chosen control variate one can also reduce the variance of measured statistics to a significantly larger degree than what is seen with paired and fixed simulations, particularly beyond the linear regime.

In \cite{chartier2020,chartier2021,Chartier:2022kjz,Ding:2022ydj}, approximate or particle-mesh $N$-body simulations were used as control variates for higher resolution $N$-body simulations. In this case, the dominant cost of the control variate method was estimating the mean of the control variate, which was still performed via Monte Carlo. In the case of \cite{Ding:2022ydj}, who used \texttt{FastPM} simulations \cite{Feng2016} for their control variate, estimation of the mean required 500 \texttt{FastPM} simulations, using approximately 24 million NERSC CPU-hours. This choice of control variate, while having the beneficial properties mentioned above, is more expensive than the method of pairing and fixing. 

This additional expense can be circumvented if an alternative control variate is chosen, such that it has an analytically calculable mean. In \cite{Kokron22}, we investigated the use of just such an alternative: the Zel'dovich approximation (ZA) \cite{1970A&A.....5...84Z,White:2014gfa}. This choice, which we call Zel'dovich control variates (ZCV), comes with the benefit of a mean prediction that is known analytically to very high precision. Furthermore, when realized with the same initial conditions, ZA matter density fields are significantly correlated with their full $N$-body counterparts out to $k\sim 1\ihmpc$. In \cite{Kokron22}, we showed that the ZA was an effective control variate for real-space matter, biased tracer auto- and cross-power spectra, and hybrid effective field theory (HEFT) basis spectra  \cite{modichenwhite19,Kokron_2021,zennaro2021bacco}. Depending on the statistic in question, ZCV produced variance reduction factors of $10^2$ to $10^6$ at $k<0.2\ihmpc$.

In this work, we further extend the ZCV method to two-point statistics in redshift space. In section~\ref{sec:controlvariates}, we give an overview of the method of control variates, and work through a simplified example demonstrating how to apply ZCV to redshift-space matter power spectra. In section~\ref{sec:za_rsd} we discuss our methodology for producing analytic and Monte Carlo ZA predictions for biased tracers in redshift space, and show that they agree with each other. Section~\ref{sec:sims} describes the simulations that we use in this work, and section~\ref{sec:results} summarizes our main results. We conclude in section~\ref{sec:conclusions}, with technical details related to our analytic predictions and estimators relegated to appendices.

\section{Control variates: a worked example}
\label{sec:controlvariates}

In this section we provide a brief overview of the steps required to reduce the variance of the redshift-space matter power spectrum, $\hat{P}_{\textrm{mm},\ell}(k)$, using the ZA as a control variate (throughout this work, we will make use of $\hat{P}$ to denote measured power spectra). In the following sections, we will describe each step in greater detail and investigate corrections beyond these steps, and their impact on the resulting variance reduction.

Control variates are a general technique that can be applied to reduce the variance of a random variable, $X$, when a correlated random variable, $C$, with known mean, $\mu_c$, is available. In this case, a new variable, $Y$, can be defined as
\begin{align*}
    Y = X - \beta (C - \mu_c).
\end{align*}
\noindent 
This is an unbiased estimator of $X$ regardless of what $\beta$ is set to, as $\langle C - \mu_c \rangle=0$ (some caveats are discussed in appendix \ref{sec:beta}). If we wish to minimize the variance of $Y$, it can be shown that the optimal choice for $\beta$ is
\begin{align}\label{eq:beta}
\beta^{\star}=\frac{\textrm{Cov}[X,C]}{\textrm{Var}[C]}.
\end{align}
\noindent 
In this case, the variance of $Y$ is given by
\begin{align*}
    \textrm{Var}[Y] &= \textrm{Var}[X](1 - \frac{\textrm{Cov}[X,C]}{\textrm{Var}[C]\textrm{Var}[X]} + \beta^{2}\textrm{Var}[\mu_c]) \\
    &= \textrm{Var}[X](1 - \rho_{xc}^2 + \beta^2\textrm{Var}[\mu_c]).
\end{align*}
\noindent
where $\rho_{xc}$ is the cross-correlation coefficient between $X$ and $C$, and $\textrm{Var}[\mu_c]$ is the uncertainty on the mean of $C$. The first line holds so long as $\mu_c$ and $\beta$ are not correlated with each other, nor with $X$. Typically, what has been done in the control variate literature in the context of cosmology is to estimate $\mu_c$ via Monte Carlo realizations of $C$, in which case the dominant cost of the control variate technique is this process. On the other hand, if $\mu_c$ is known analytically, then this additional expense can be circumvented, and $\textrm{Var}[Y] = \textrm{Var}[X](1 - \rho_{xc}^2)$. 

In cosmology, $X$ is often a measurement, such as the redshift-space matter power spectrum, $\hat{P}_{\rm{mm}, \ell}(k)$, estimated from an expensive simulation. In this case, we would like to choose $C$ such that it is highly correlated with the redshift-space matter field, while still having an analytically known mean. In \cite{Kokron22}, we showed that the ZA satisfies both of these requirements for real space fields, and in this work, we demonstrate that this also holds in redshift space, for matter, halos and simulated galaxies. 

The algorithm for using the ZA to reduce the variance of $\hat{P}_{\rm{mm}, \ell}(k)$ measured from an $N$-body simulation is as follows:
\begin{enumerate}
    \item Generate an initial linear density field $\delta(\mathbf{k})$ using the same random seed as used to run the $N$-body simulation, and use this to compute Zel'dovich displacements, $\Psi(\mathbf{q})$.
    \item Compute the redshift-space Zel'dovich density field $\delta_s^{\rm ZA}(\mathbf{k},z)$ at the desired redshift $z$ using the linear growth factor, $D(z)$, and growth rate, $f(z)$.
    \item Measure the ZA and $N$-body matter auto-power spectra, and ZA--$N$-body cross-power spectra, $\hat{P}^{zz}_{\rm{mm},\ell}(k), \hat{P}^{nn}_{\rm{mm},\ell}(k)$, and $\hat{P}^{zn}_{\rm{mm},\ell}(k)$ and estimate $\beta$ using a disconnected approximation for $\textrm{Cov}[\hat{P}^{zz}_{\rm{mm},\ell}(k), \hat{P}^{nn}_{\rm{mm},\ell}(k)]$.
    \item Analytically compute the mean Zel'dovich redshift-space auto-power spectrum, $P^{zz}_{\rm{mm}, \ell}(k)$, and construct the reduced variance matter power spectrum via $\hat{P}^{\ast, nn}_{\rm{mm},\ell}(k) = \hat{P}^{nn}_{\rm{mm},\ell}(k) - \beta\left(\hat{P}^{zz}_{\rm{mm},\ell}(k) - P^{zz}_{\rm{mm},\ell}(k)\right)$.
    
\end{enumerate}

\begin{figure}[t]
\centering
      \includegraphics[width=\columnwidth]{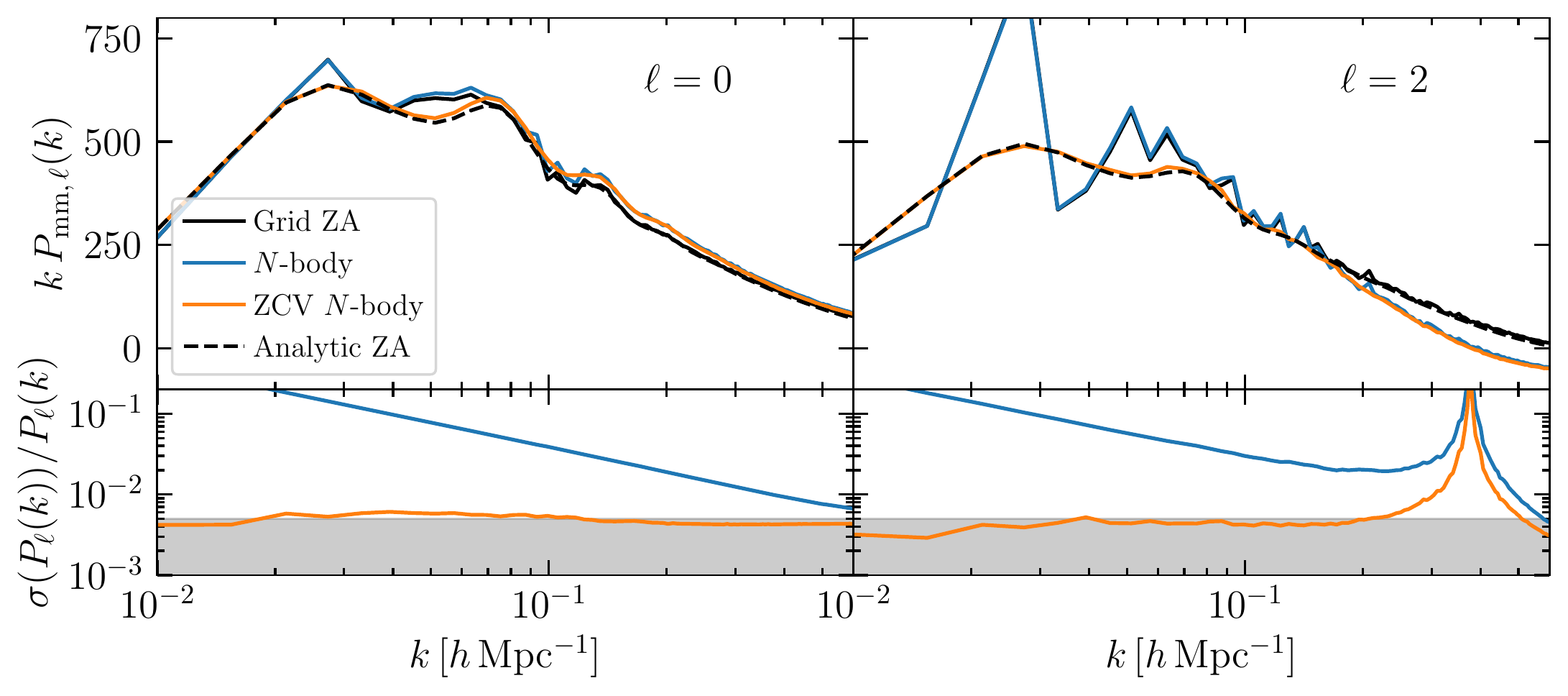}
      \caption{(\textit{Top}) Comparison of redshift-space matter power spectrum multipoles measured from our fiducial $(1050\, h^{-1}\, \rm{Mpc})^3$ $N$-body simulation (blue), a numerical realization of the ZA using the same initial linear density field as the $N$-body simulation (black), and the $N$-body spectrum with reduced variance using the black curve as a control variate (orange). For reference, we also display the mean prediction for the Zel'dovich matter power spectrum (black dashed). On large scales, the variance reduced $N$-body spectra agree nearly perfectly with the noiseless analytic ZA prediction. (\textit{Bottom}) Fractional errors, computed via a disconnected covariance approximation (blue), and the variance reduced version of these errors (orange). The variance reduction is given by $1-\rho_{xc}^2$. The shaded gray region below $0.5\%$ fractional error is approximately where we expect the contribution from systematic errors in simulations to become appreciable.}\label{fig:pmm_example}
\end{figure}

The results of this process are summarized in figure~\ref{fig:pmm_example}. The blue curves in the top panels display the monopole and quadrupole moments of the redshift-space matter power spectrum measured from our $N$-body simulation, with their fractional errors, estimated using a disconnected covariance approximation, shown in blue in the bottom panels. In solid black, we also display the ZA matter power spectra produced with the same initial conditions. On large scales, these curves are almost indistinguishable, indicating the large amount of correlation that exists between the ZA and $N$-body matter fields in redshift space. The black dashed line displays the analytic prediction for $P^{zz}_{\rm{mm},\ell}(k)$, whose computation we will briefly overview in \S\ref{sec:za_rsd}. The orange curve in the bottom panel displays the fractional error on $\hat{P}^{\ast,nn}_{\rm{mm},\ell}(k)$, which is reduced with respect to the blue curve by a factor of $1-\rho_{xc}^2$. $\rho_{xc}$ is estimated from a single simulation as described in section~\ref{sec:biased_tracers} and appendix \ref{sec:beta}. We see that for the redshift-space matter power spectrum we are able to reduce the errors on our measurements from a $(1050\, h^{-1}\, \rm{Mpc})^3$ simulation to yield fractional errors that are at or below the $\sim 0.5\%$ expected contribution from simulation systematic errors, represented by the shaded gray regions in the bottom panel \cite{Schneider_2016,Garrison2021b, grove2022}.

\begin{figure}
\centering
      \includegraphics[width=\columnwidth]{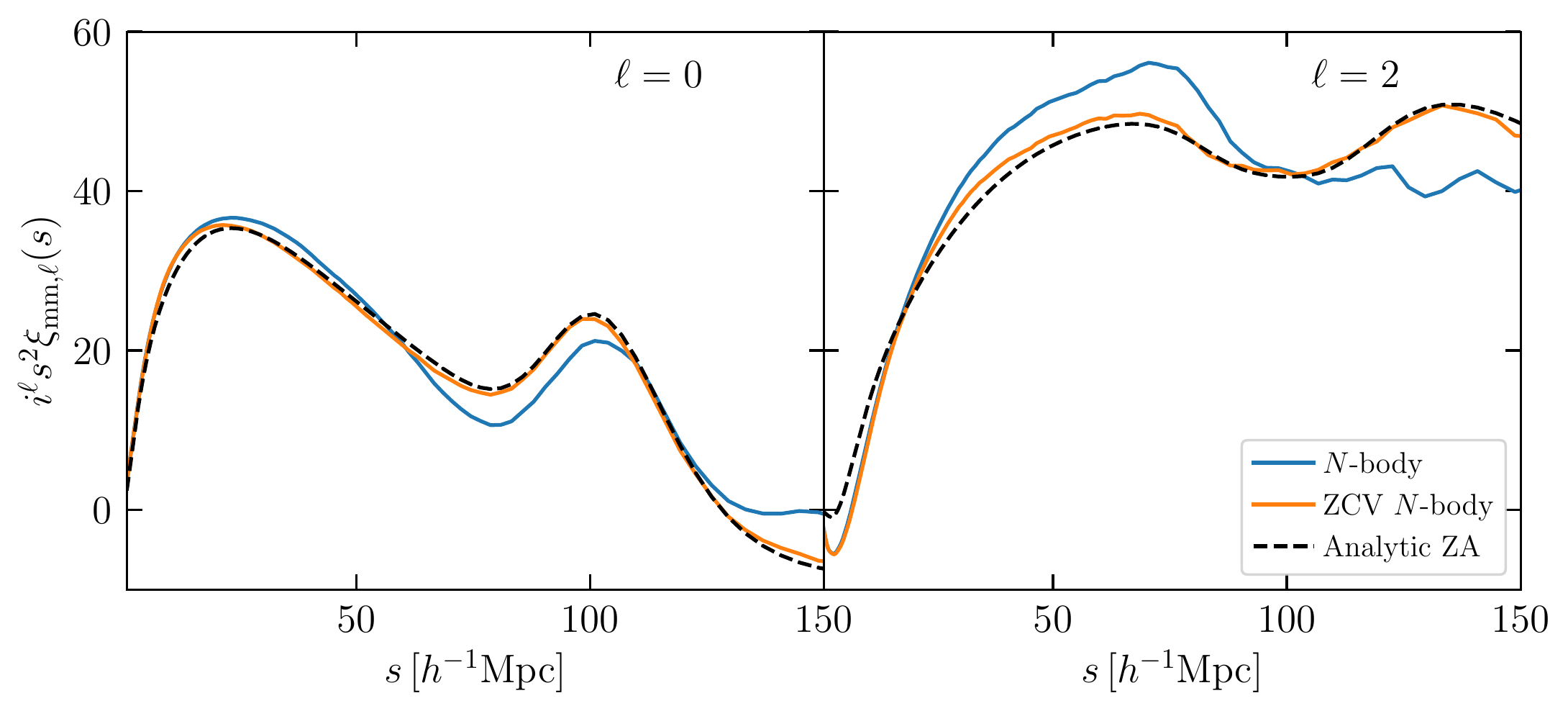}
      \caption{Matter correlation function multipoles with (orange) and without (blue) applying the ZCV method, compared to our analytic ZA theory (dashed). We have Hankel transformed the measured power spectra in order to produce correlation functions as described in the text. A significant reduction in noise is apparent in the correlation functions that have ZCV applied.}
      \label{fig:corr_func}
\end{figure}

For the purposes of demonstration, we also display the effect of applying the ZCV method to correlation functions in figure~\ref{fig:corr_func}. Here, we apply Hankel transforms to the redshift-space matter power spectra in figure~\ref{fig:pmm_example}, extrapolating at $k<0.01\ihmpc$ using our analytic ZA model, and at $k>4$ using a power law extrapolation. The comparison between the correlation functions with and without ZCV applied is quite stark, with the large scales matching ZA nearly perfectly in the ZCV case, while significant noise is apparent in its absence. Correlation functions can be estimated more accurately by using ZCV to reduce the variance of the un-binned three-dimensional redshift-space power spectrum, Fourier transforming to three-dimensional correlation functions, and measuring multipoles. This would avoid any reliance on interpolation and extrapolation. Doing so is readily possible with the methods presented in this work.

In the following sections, we will elaborate on this example, demonstrating the efficacy of ZCV for redshift-space galaxy and halo power spectra. In order to do this, we will introduce bias operators in our Zel'dovich field in section~\ref{sec:biased_tracers}. We have also glossed over a number of technical details that are important for obtaining an unbiased control variate estimator, such as window function deconvolution and the estimation of $\beta$ that are discussed in detail in Appendix \ref{sec:window} and \ref{sec:beta}, respectively.

\section{The Zel'dovich approximation in redshift space}
\label{sec:za_rsd}

Within a cold dark matter universe, structure formation is entirely determined by the displacements $\Psi(\bq,a)$, or trajectories, of phase-space elements over cosmic time. The displacements translate between the initial positions $\bq$ of these elements and their positions $\bx = \bq + \Psi(\bq,a)$ at scale factor $a$. The matter density $\delta_m$ is obtained by counting the particles that end up at position $\bx$ at $a$, i.e.
\begin{equation}
    1 + \delta_m(\bx,a) = \int d^3\bq \ F(\bq)\ \delta_D(\bx - \bq - \Psi(\bq,a)),
    \label{eqn:density_field}
\end{equation}
where the weight given to each dark matter particle is $F(\bq) = 1$, so that the non-linear clustering of matter can be entirely written in terms of the statistics of the displacements. For biased tracers like galaxies we need to additionally weight each $\bq$ by functions of the initial conditions using the so-called bias expansion (see e.g.\ \cite{Matsubara2008,Desjacques:2016bnm}) such that
\begin{equation}
    F(\bq) = 1 + b_1 \delta_0 + b_2 \big( \delta_0(\bq)^2 - \langle \delta_0^2 \rangle \big) + b_s \big( s_0(\bq)^2 - \langle s_0^2 \rangle \big) + ...
    \label{eqn:bias_expansion}
\end{equation}
where $s^2 = s_{ij} s_{ij}$ is the local shear field. From eq. \ref{eqn:density_field} we can then see that the final biased tracer field, $\delta_t(\bx,a)$, can be expressed in terms of an expansion in the advected initial fields, 
\begin{align}
    1 + \delta_t(\bx,a) = \sum_{\mo_i \in m, \delta_0,\, \delta_0^2,\, s_0^2,\, \ldots} b_{\mo_i}\mathcal{O}_i(\bx,a)\, ,
    \label{eq:advection}
\end{align}
where the advected operators $O_i$ are given in both configuration and Fourier space by
\begin{align}
    \mathcal{O}_i(\bx,a) &= \int d^3\bq \ \mathcal{O}_i(\bq)\ \delta_D(\bx - \bq - \Psi(\bq,a)) \nonumber \\
    \mathcal{O}_i(\bk,a) &= \int d^3\bq \ e^{-i\bk\cdot(\bq+\Psi(\bq))}  \ \mathcal{O}_i(\bq)
    \label{eq:zel_ops}
\end{align}
\noindent 
where we will denote the advected initial fields as $\mathcal{O}_i(\bx,a)$. Furthermore, for notational convenience, we have set $\mathcal{O}_m(\bq)=1$, $b_{\mo_m}=1$, and thus $\mathcal{O}_m(\bx,a)=\delta_m(\bx,a)$.

There exist many ways to compute the displacements, $\Psi$. Perhaps the most straightforward is to directly simulate the dynamics of structure formation using $N$-body simulations. In this case, the Lagrangian position $\bq$ is the grid position set in the initial conditions, and $\Psi$ is simply given by the trajectory of the $N$-body particle over the simulated time. Within $\Lambda$CDM universes close to our own, however, the largest contribution to $\Psi$ comes from large-scale bulk flows that are extremely well captured by first-order Lagrangian perturbation theory (LPT), also known as the Zel'dovich approximation (ZA) \cite{1970A&A.....5...84Z}. The advection of dark matter particles away from their initial positions $\bq$ by the Zel'dovich displacement is the leading contribution to the decorrelation of large scale structure with the initial conditions \cite{Chisari19}, and the cancellation of this effect by using Zel'dovich power spectra as our control variate is the main advantage of ZCV over linear initial condition control variates, as we show in \S\ref{sec:linear_theory}.

In the ZA, the displacement is set by the initial gravitational potential gradient experienced by the particle, given by Poisson's equation to be $k^2 \Phi \propto \delta_m$, such that 
\begin{equation}
    \Psi^{(1)}(\bk,a) = \frac{i \bk}{k^2}\  D(a)\ \delta_0(\bk)
    \label{eq:za_displ}
\end{equation}
where $D(a)$ is the linear-theory growth factor. Likewise, peculiar velocities are linearly proportional to $\bv^{(1)} = f \mathcal{H} \Psi^{(1)}$, where $f = d\ln D /d\ln a$ is the linear growth rate and $\mathcal{H}$ is the conformal Hubble parameter. The peculiar velocity is important because, in most situations, the distance to a galaxy is inferred from its cosmological redshift, which receives a contribution from the peculiar line-of-sight velocity $\textbf{u} = \bv_{\parallel} / \mathcal{H}$. Within the ZA this is equivalent to multiplying the displacement by a constant matrix \cite{Matsubara_2008}
\begin{equation}
    \Psi^{(1)}_{s, i} = R_{ij} \Psi^{(1)}_j, \quad R_{ij} = \delta_{ij} + f \hat{n}_i \hat{n}_j.
    \label{eq:zspace_za_displ}
\end{equation}
where $\hat{n}$ is the line-of-sight unit vector.

In this paper we will be mostly concerned with the redshift-space 2-point function in configuration space and Fourier space, i.e. the correlation function and power spectrum. From Equation~\ref{eqn:density_field} one can show \cite{Carlson_2012} that the biased tracer power spectrum is given by
\begin{equation}
    P_s^{tt}(\bk) = \sum_{\mo_i,\mo_j} b_{\mo_i}b_{\mo_j} P_{ij,\,s}(\bk)
    \label{eq:biased_tracer_spectra}
\end{equation}
where we have defined the \textit{basis spectra}
\begin{equation}
    P_{ij,s}(\bk) (2\pi)^3 \delta_D(\bk+\bk') = \Big \langle \mo_{i,s}(\bk)\mo_{j,s}(\bk')\Big \rangle.
\end{equation}
The operators $\mo_{i,s}(\bk)$ are the redshift-space counterparts of $\mo_{i}(\bk)$, defined such that $\Psi$ is replaced by $\Psi_s$ in their construction. We make predictions for the ensemble mean of these statistics with the  \texttt{ZeNBu} code\footnote{\href{https://github.com/sfschen/ZeNBu}{https://github.com/sfschen/ZeNBu}}.

For a given realization of the initial conditions on a grid we can numerically construct these Zel'dovich operators and their spectra by  realizing the linear density field $\delta_0$ on a cubic grid, and constructing the relevant initial condition fields, $\delta_0^2,\, s_0^2$, from $\delta_0$. We can then numerically compute the displacements using Eqs.~\ref{eq:za_displ} and \ref{eq:zspace_za_displ}, which we can use to perform the advection integral in Eq.~\ref{eq:advection}, using $\Psi(\bq)$ and $\Psi_s(\bq)$ for real and redshift-space quantities, respectively. Predicting biased tracer power spectra is then as simple as measuring the cross-power spectra between the advected operators and performing the sum in Eq.~\ref{eq:biased_tracer_spectra}. We make use of \texttt{CLASS} \cite{Lesgourgues11,Blas11} to compute linear power spectra, growth and growth rates, and \texttt{Monofonic} \cite{Michaux:2020yis} for our grid-based predictions of $\delta_0$ and $\Psi$. The code to compute the rest of the quantities required for this work is publicly available\footnote{\href{https://github.com/kokron/anzu}{https://github.com/kokron/anzu}}.

We can also compute these spectra analytically without sample variance. Within the ZA, the expectation value of the exponentiated displacements in eq.~\ref{eq:zel_ops} can be solved exactly since the displacements are purely Gaussian; for the bias expansion in Equation~\ref{eqn:bias_expansion} this was done in real space to all orders in ref.~\cite{Kokron22}. In order to perform the same calculation in redshift space, it is necessary to transform the displacements via the matrix $R_{ij}$, or alternatively by transforming the Fourier space wavenumber by the same amount, since $\bk^{T} (\textbf{R} \Delta) = (\bk \textbf{R})^T \Delta.$ Numerical methods to compute the quantities in the latter choice of coordinates were developed in \cite{Taylor82,Vlah19,Chen19,chen2020redshiftspace}; roughly, the azimuthally symmetric real space integrals are replaced with ones accounting for the line-of-sight dependence in redshift space. We review these developments in the specific context of the ZA in Appendix~\ref{app:zenbu}.

\begin{figure}
\centering
      \includegraphics[width=\columnwidth]{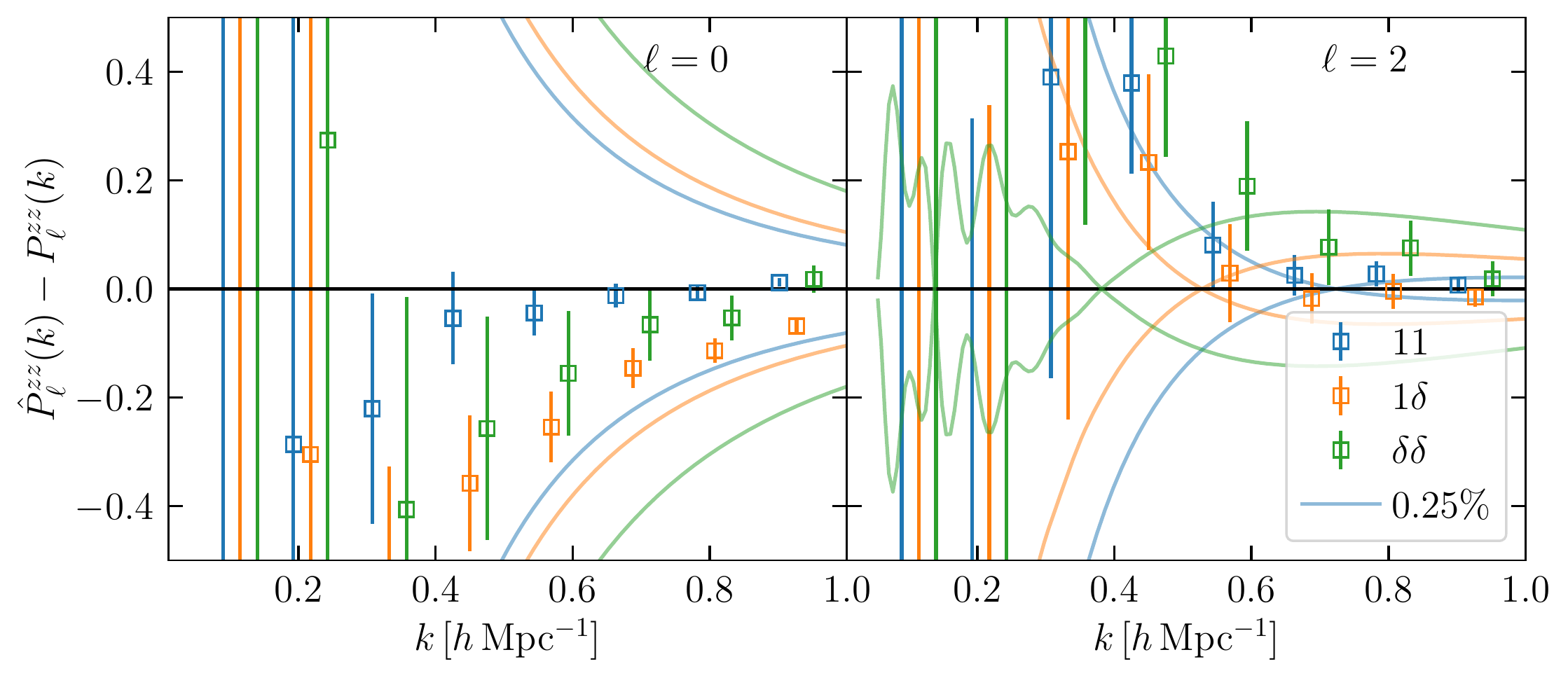}
      \caption{The difference between our grid-based ZA realizations and their analytic predictions for $P_{ij,\ell}(k)$ up to linear order in $\delta_0$. These measurements are averaged over 60 random seeds, and coarsely binned in $k$ to suppress variance. We have offset the measurements for the different basis spectra in $k$ to aid in presentation. The lines displayed are $0.0025 P_{ij,\ell}(k)$, i.e. they represent a $0.25\%$ error in each spectrum. Away from zero-crossings, the difference between our grid-based and analytic ensemble mean predictions is consistent with or less than this, and thus any uncertainty due to our ability to predict the mean of $ P_{ij,\ell}(k)$ is negligible. We find consistent behavior for all other basis spectra, but we do not display them in order to facilitate presentation. On large scales, a proper treatment of the mixing of multipoles due to discreteness effects in our simulation is of great importance for obtaining this level of agreement, see app.~\ref{sec:window}.}
      \label{fig:zenbu_comp}
\end{figure}

The validity of the Zel'dovich control variate method depends on our ability to very accurately predict the ensemble average of the grid-based realizations, averaging over realizations of $\delta_0$. In \cite{Kokron22}, we showed that we were able to do this in real space after smoothing $\delta_0$ using a Gaussian kernel with a width equal to the mesh scale, $N\pi/L$ where $N$ is the mesh size and $L$ is the side length of the simulation cube. We find that we are able to predict our redshift-space grid-based results with our analytic code to comparable accuracy after performing this same smoothing procedure, using two times the smoothing scale, i.e.\ $k_{\rm smooth}=N\pi/(2L)$, as shown in figure~\ref{fig:zenbu_comp}. We were not able to achieve a comparable level of accuracy for the hexadecapole, and so do not treat it in this work. Care must be taken to correct for artifacts in our grid-based measurements imparted due to discrete $\bk$ sampling in our simulations. This is discussed in Appendix~\ref{sec:window}.

\section{Simulations and galaxy models}\label{sec:sims}

In this work, we employ the Aemulus $\nu$ simulations \cite{DeRose2022c}, a suite of CDM and massive neutrino simulations run in a similar configuration to the Aemulus $\alpha$ simulations \cite{DeRose2018}, but with an expanded cosmological parameter space, and a greater number of realizations. Each simulation has a volume of $(1.05 h^{-1}\rm{Gpc})^3$, evolving $1400^3$ CDM particles, and the same number of neutrino particles. The simulations are initialized at $z=12$ with 3rd-order Lagrangian perturbation theory, using a version of the \texttt{Monofonic} code \cite{Michaux:2020yis} modified to account for the presence of neutrinos on the evolution of $\delta_{\rm cb}$ \cite{Elbers2022}. Further details and convergence tests of these simulations will be presented in \cite{DeRose2022c}. In this work, we make use of the $\alpha\texttt{000}$ simulation with a cosmology of $n_s=0.97$, $H_{0}=67.0$, $w=-1.0$, $\omega_b=0.0223$, $\omega_{c}=0.12$, $\sum M_{\nu}=0.07\rm eV$, $\sigma_8=0.81$. For simplicity, we will conflate matter and $\delta_{cb}$ power spectra in this work, as the difference between the two is unimportant for all of the findings presented here. Halo finding was performed using the \texttt{ROCKSTAR} spherical overdensity halo finder \cite{Behroozi_2013} and $M_{200b}$ strict spherical overdensity (SO) masses. 

When populating galaxies in our simulation, we make use of the halo occupation distribution (HOD) formalism \citep{Seljak2000,Berlind_2002}. In particular, we present tests using galaxies populated with HODs consistent with the DESI luminous red galaxy (LRG) sample, making use of the HOD functional form presented in \cite{Zheng:2007zg}. We take as parameters of this model the best-fit values for the second redshift bin presented in \cite{Zhou_2020}. For the LRG sample, we make use of the $z=0.56$ output from our simulation, yielding a number density of $\bar{n}=3\times10^{-4}\, h^{3}\rm Mpc^{-3}$.

\section{Results and discussion}
\label{sec:results}
Here we demonstrate the effectiveness of ZCV for halo and galaxy samples, and discuss some of the complications that arise in these cases. We begin by investigating the impact of biasing, followed by satellite galaxies and their accompanying addition of stochastic small scale velocities. We will see that even in the presence of these complications, we can reduce the sample variance of redshift-space power spectra down to the shot-noise limit until $k\sim 0.2\ihmpc$. We then compare the ZCV method to a simpler method using linear theory as a control variate. Finally, we take advantage of the demonstrated performance of ZCV to forecast the amount of variance reduction we expect to achieve for a number of key galaxy samples that will be used for cosmological studies in the coming years. For most of the samples we consider, we find variance reduction factors of $\sim10-100\times$.

\subsection{Biased tracers}
\label{sec:biased_tracers}

In section~\ref{sec:controlvariates}, we introduced the ZCV method in the context of the redshift-space matter power spectrum. While this application may be useful in some contexts, it represents an idealized case. Instead, one is often interested in measuring power spectra of biased tracers such as halos or galaxies. In this section, we will investigate the extent to which ZA predictions are correlated with these statistics. Importantly, we wish to disentangle the effects that biasing, shot noise, and stochastic small scale velocities, i.e.\  the finger of god (FoG), have on the efficacy of the ZCV method. In order to do so, we will first consider ZCV in the context of power spectra of halos in bins of mass. By varying halo mass, we aim to isolate the impact that halo bias, both linear and non-linear, as well as shot noise, has on the performance of the ZCV algorithm.

First, we consider the ability of the Zel'dovich matter power spectrum to act as a control variate for the power spectra of halo samples split into three logarithmically spaced bins in mass between $10^{13}\hmsun$ and $10^{14.5}\hmsun$.  We have chosen relatively high mass halos to emphasize the role of shot noise, which leads to decorrelation and hence reduces the effectiveness of the control variates technique. In order to quantify the effectiveness of the ZA as a control variate, we measure the cross correlation coefficient, $\rho_{xc}$, between the redshift-space power spectra of these halo samples, and the redshift-space Zel'dovich matter power spectrum. We compute this quantity as
\begin{align}
    \rho_{xc} &= \frac{\textrm{Cov}[\hat{P}^{tt}_{\ell}(k), \hat{P}^{zz}_{\ell}(k)]}{\sqrt{\textrm{Var}[\hat{P}^{tt}_{\ell}(k)]\textrm{Var}[\hat{P}^{zz}_{\ell}(k)]}}\,
    \label{eq:rhoxc}
\end{align}
\noindent 
where $\hat{P}^{tt}_{\ell}(k)$ are the multipoles of the redshift-space power spectrum of the (biased) tracer under consideration, and $\hat{P}^{zz}_{\ell}(k)$ are the same for the ZA field that we employ as a control variate. The fractional variance reduction afforded by a particular control variate is equal to $1-\rho_{xc}^2$, making this the main quantity of interest in our study. We approximate the covariances and variances in eq.~\ref{eq:rhoxc}, keeping only their disconnected contributions, as described in app.\ \ref{sec:beta}. In this case, the Zel'dovich field is simply the unweighted matter field, but we will make use of the same method for measuring $\rho_{xc}$ when we include bias operators in the ZA. We expect the disconnected approximation to break down at high $k$, but for $k<0.2\ihmpc$ it holds to very high accuracy \cite{wadekar2020}. As we will see, above $k\sim0.2\ihmpc$ the ZA fields decorrelate from the fields of interest, and so break-downs in this approximation are not important for the conclusions of this work. 

\begin{figure}
\centering
      \includegraphics[width=\columnwidth]{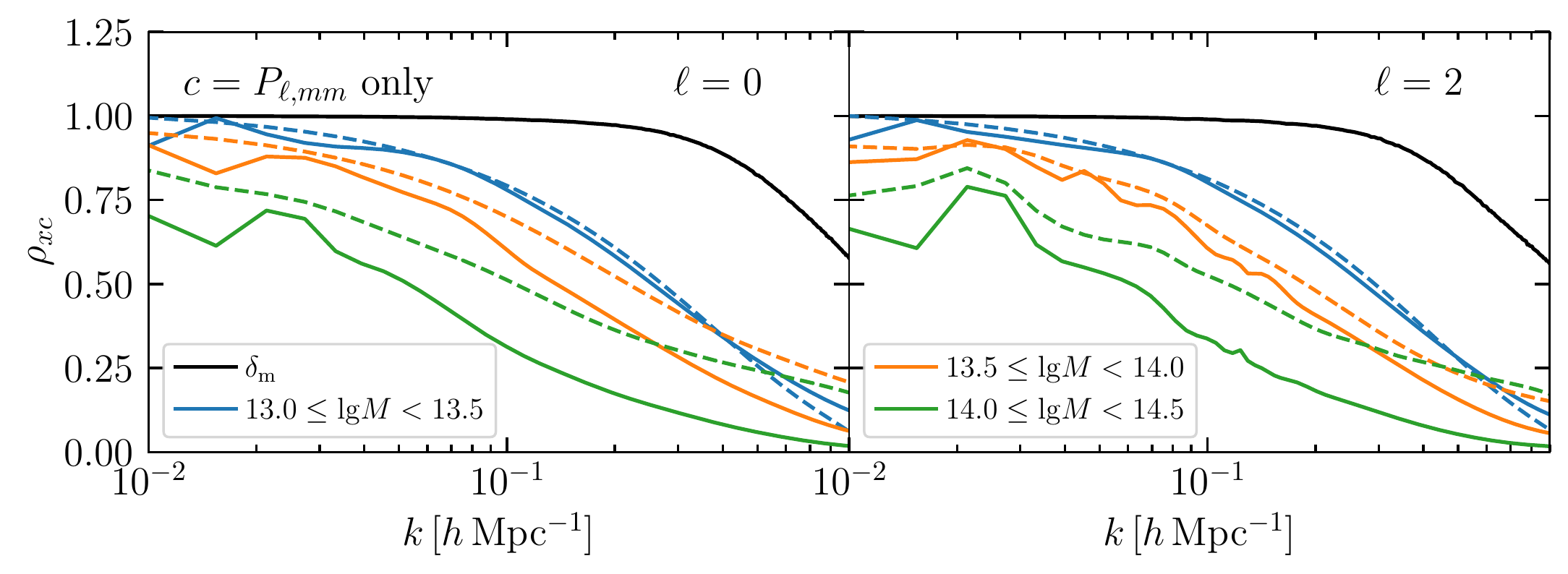}
      \caption{Cross-correlation coefficient between the redshift-space ZA \textit{matter} power spectrum and various tracer power spectra, including the non-linear matter field, and halos in three bins of mass as denoted in the legend. The different panels show the cross correlation between different multipoles of the redshift-space power spectrum. The dashed lines are the cross-correlation coefficient that we would expect if the tracer field were perfectly correlated with the ZA field apart from a constant shot-noise contribution. The different multipoles exhibit very similar behavior, with more biased, higher shot-noise samples decorrelating more. As discussed later, including higher-order bias operators allows us to saturate the shot-noise limit for all samples to $k\sim0.2\ihmpc$.}
      \label{fig:halo_rho_xc_nobias}
\end{figure}

Figure \ref{fig:halo_rho_xc_nobias} displays $\rho_{xc}$ for each of the three halo mass bins that we consider. For reference, we also compare $\rho_{xc}$ between the nonlinear redshift-space matter power spectrum, $\hat{P}_{\rm mm,\ell}(k)$, and the Zeldovich redshift-space matter power spectrum, $\hat{P}^{zz}_{\rm mm,\ell}(k)$, in black. The main aspect of note is that $\rho_{xc}$ decreases monotonically with halo mass, and this decrease becomes even more pronounced with respect to the $\hat{P}_{\rm mm,\ell}$ case at high $k$. The dashed lines in this figure show the expectation for $\rho_{xc}$ in the limiting case where the halo power spectra correlate perfectly with their ZA counterparts except for the shot-noise contribution to the former (we describe how we estimate the shot noise momentarily). These dashed lines are the best one can do without incorporating a control variate with the same shot-noise properties as the tracer sample in question. Finding such a control variate is highly nontrivial, as is analytically or cheaply computing its mean, as the shot-noise level depends on a number of things other than the number density of the sample in question, such as halo exclusion, and non-linear biasing  effects \cite{Baldauf_2013, Hamaus_2010, Kokron:2021faa}.

Now we proceed to investigate the extent to which including bias operators in our Zel'dovich field increases $\rho_{xc}$ for the halo power spectra discussed above. In order to include additional bias operators in the ZA, we must fit for the bias coefficients of each tracer sample. We do this at the field level, minimizing the residuals between the Zel'dovich and halo fields in real space by minimizing
\begin{align*}
    S &= \langle \epsilon(\bx)^2 \rangle_{|\bk|<k_{\rm max}} \\
      & \approx \int_{|\bk|<k_{\rm max}} \frac{d^3\bk}{(2\pi)^3} \bigg| \delta^{t}(\bk) - \delta^{z}_{m}(\bk) - \sum_{\mo_i} b_{\mo_i}\mathcal{O}_i(\bk) \bigg|^2\, ,
\end{align*}
following the procedure outlined in section 2.3 of \cite{Kokron:2021faa}, but replacing the HEFT operators with their ZA counterparts. We perform fits including the $\delta_0(\bk),\,\delta_0^2(\bk)$ and $s_0^2(\bk)$ fields. This procedure also allows us to precisely determine the shot-noise contribution to the tracer power spectra, because at low-$k$, shot noise is equivalent to the error power spectrum:
\begin{align*}
P^{\epsilon\epsilon}(k) &= \langle \epsilon(\mathbf{k}) \epsilon(-\mathbf{k}) \rangle \\
                        &= \langle [\delta^{t}(\mathbf{k})-\delta^{z}(\mathbf{k})] [\delta^{t}(\mathbf{-k})-\delta^{z}(\mathbf{-k})] \rangle\, ,
\end{align*}
\noindent 
where we have applied the shorthand $\delta^{z}(\bk)=\delta^{z}_{m}(\bk) + \sum_{\mo_i}b_{\mo_i}\mathcal{O}_i(\bk)$. 

We perform fits to $k_{\rm max}=0.15\ihmpc$, and average $\hat{P}^{\epsilon\epsilon}(k)$ over this range in order to estimate the shot-noise values. We have confirmed that our shot-noise estimates do not change when including the effects of nonlinear displacements, e.g. by replacing the Zel'dovich fields used above with their HEFT counterparts. We make use of this estimate of shot noise, rather than the Poissonian expectation, because it incorporates the effect of higher-order biasing and halo exclusion \cite{Kokron:2021faa}. Poissonian shot-noise estimates would mostly overestimate the impact of shot noise on the performance of ZCV for our halo and galaxy samples, as halo exclusion effects tend to reduce the level of shot noise compared to the Poisson expectation for the halo mass range used here, leading to lower levels of uncorrelated noise than the naive estimate.

With an estimate of the shot noise of the tracer in question, we can then construct our estimate of the shot-noise limit of $\rho_{xc}$ as
\begin{align}
    \rho_{xc}^{\rm SN.\, limit} &= \frac{\textrm{Var}_{\rm no SN}[\hat{P}^{tt}_{\ell}(k)]}{\sqrt{\textrm{Var}[\hat{P}^{tt}_{\ell}(k)]\textrm{Var}_{\rm no SN}[\hat{P}^{tt}_{\ell}(k)}]}\,
    \label{eq:rhoxc_snlimit}
\end{align}
where $\textrm{Var}_{\rm noSN}[\hat{P}^{tt}_{\ell}(k)]$ is estimated by subtracting $\langle \hat{P}^{\epsilon\epsilon}(\bk)\rangle_{|\bk|<k_{\rm max}}$ from the monopole in the disconnected approximation for the variance.

\begin{figure}
\centering
      \includegraphics[width=\columnwidth]{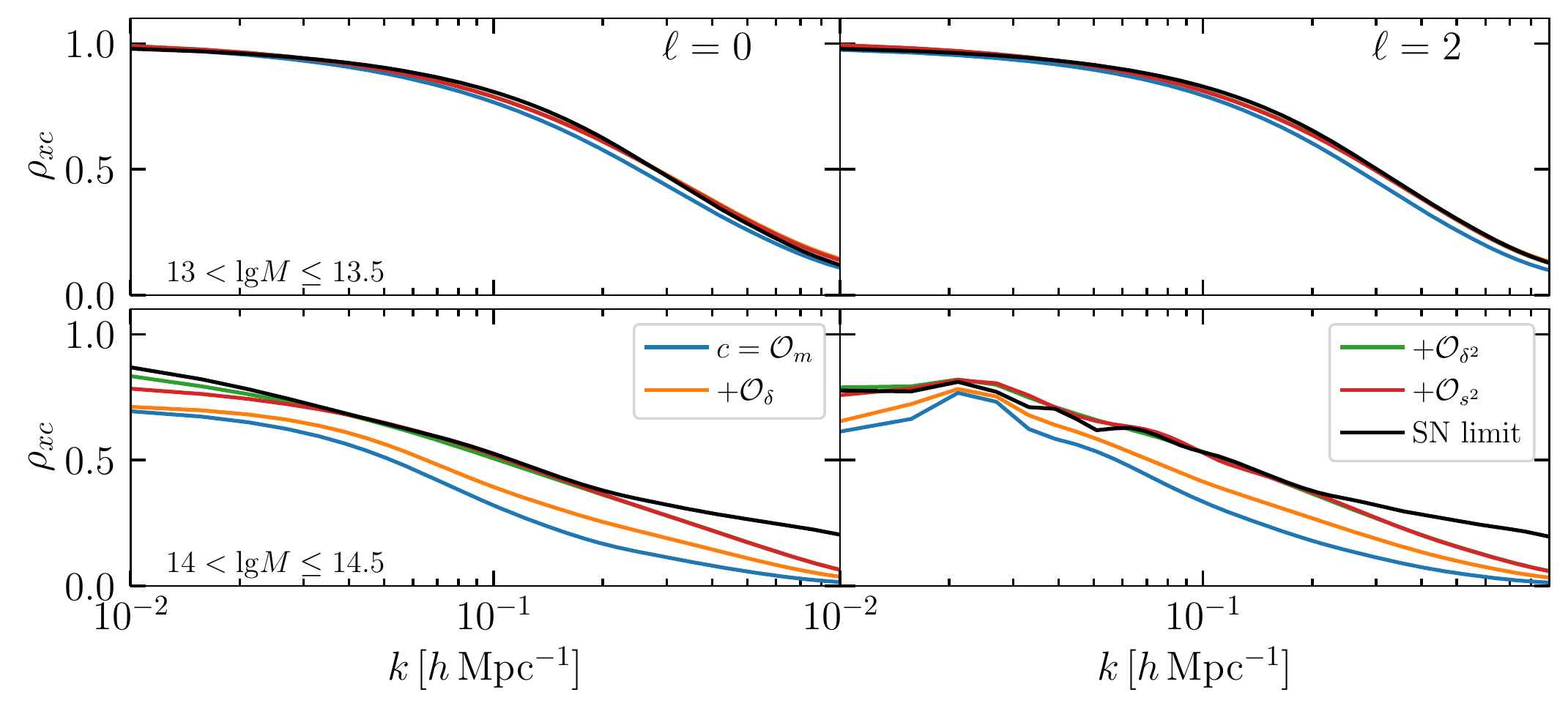}
      \caption{The effect of including additional bias terms in the ZA on the cross-correlation coefficient between multipoles of the redshift-space Zel'dovich and halo power spectra.  We show the two extreme halo mass bins: $13<\textrm{lg}M\le 13.5$ (top) and $14<\textrm{lg}M\le 14.5$ (bottom). For reference, we display the expectation for the correlation coefficient if all of the signal in the field were perfectly correlated (and only shot noise contributed to decorrelation) in solid black. We saturate the shot-noise limit in both bins, although including higher order bias operators is much more important in doing so for the higher mass sample. 
      } 
      \label{fig:halo_rho_xc_biasscan}
\end{figure}

Figure \ref{fig:halo_rho_xc_biasscan} demonstrates the extent to which adding bias operators to our ZA control variate can optimize the sample variance reduction properties of the ZCV method. The top and bottom panels demonstrate two different regimes. The top shows the behavior of ZCV for a relatively low-bias halo sample, with bias values of $b_{\delta_0}=1.06$, $b_{\delta_0^2}=-0.14$ and $b_{s_0^2}=0.26$. We see that including additional bias operators makes a relatively minor difference. Using the unbiased (i.e.\ matter) ZA redshift-space power spectrum nearly saturates the shot-noise limit, and operators beyond linear bias yield no additional improvement. 

The bottom panel displays a much higher bias halo sample, with $b_{\delta_0}=3.72$, $b_{\delta_0^2}=9.09$ and $b_{s_0^2}=-3.0$. Here we see large improvements going from the unbiased ZA, to linear bias and quadratic bias, with the latter saturating the shot-noise limit to $k\sim0.2\ihmpc$. Somewhat unsurprisingly, we never observe improvements when adding in tidal bias. This is simply explained by the fact that the basis spectra, $P_{ij}(k)$, that contain the tidal field operator in eq.~\ref{eq:biased_tracer_spectra} are highly subdominant to the linear and quadratic bias terms on scales where the ZA is still correlated with our halo fields. As such, it seems unlikely that tidal bias operators will ever be required for an optimal ZCV implementation.

\subsection{Satellite galaxies}

Having seen that ZCV is able to saturate the shot-noise limit for halos, we now turn to investigating whether this still holds for galaxy samples populated via HODs. In particular, one might expect the addition of satellite galaxies to boost stochastic contributions from non-perturbative small scale velocities, i.e.\ the FoG, thus decorrelating our ZA control variate from the galaxy field more than what we found for halos. We consider three different samples for this test. First, we examine the LRG-like HOD described in Sec.~\ref{sec:sims}, taken from \cite{Zhou_2020}. We also consider a galaxy sample with two times the satellite fraction of the LRG-like sample, i.e.\  $f_{\rm sat}=0.3$ compared to the original $f_{\rm sat}=0.15$, which we have achieved by setting $\textrm{lg} M_1=13.6$ and leaving the remaining HOD parameters fixed. Finally, we use a sample containing only the central galaxies from the LRG-like HOD, thus minimizing the FoG effect but otherwise keeping the mean halo mass nearly the same.

\begin{figure}
\centering
      \includegraphics[width=\columnwidth]{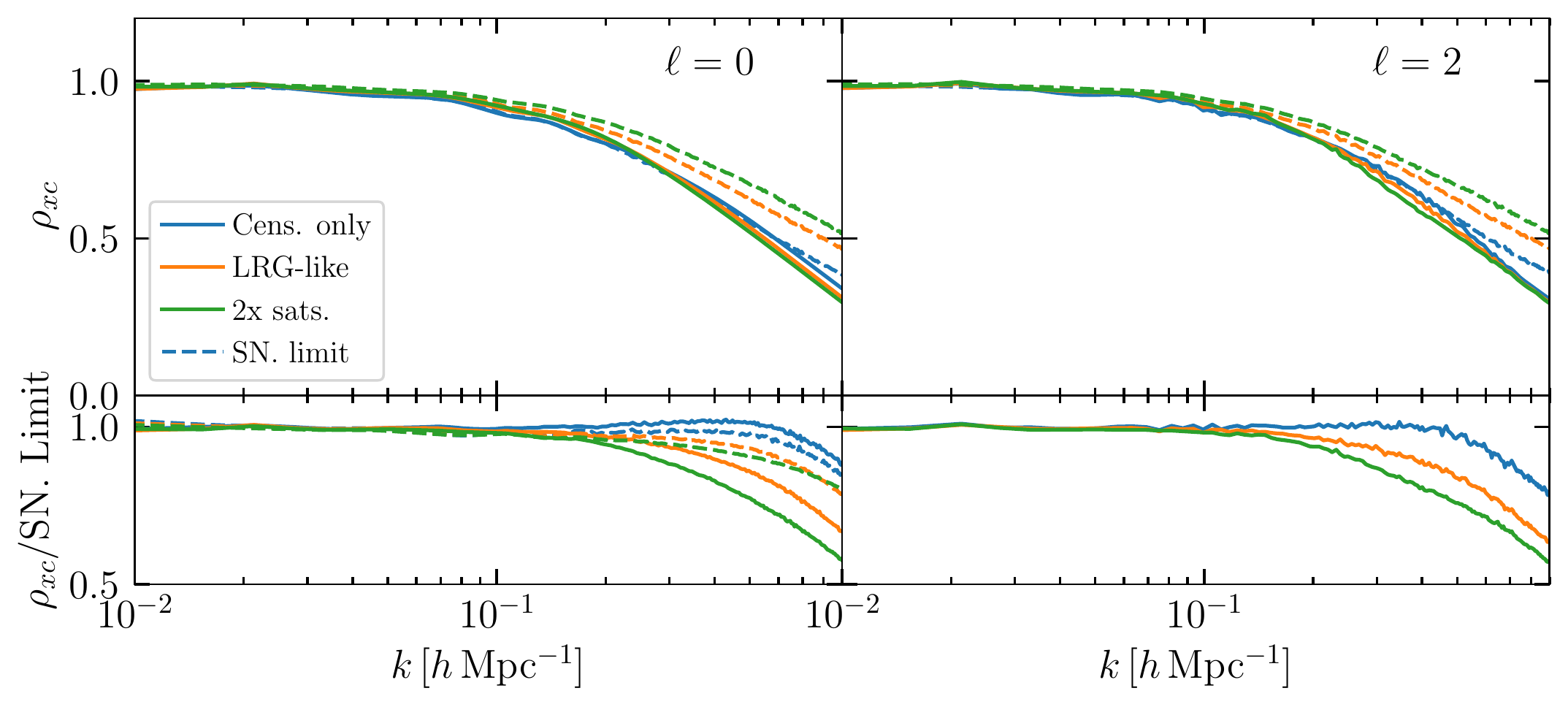}
      \caption{(\textit{Top}) Comparison of $\rho_{xc}$ for an LRG-like galaxy sample, the central galaxies from that sample, and an HOD with two times the number of satellite galaxies as the fiducial HOD. All samples use a control variate that includes linear and quadratic bias. The shot-noise limit for each sample is shown by the dashed lines. To $k\sim0.2\ihmpc$, we find that we are able to reduce the variance of all the samples to this limit. At higher $k$, the extent to which we are able to reduce the variance is sensitive to decorrelation between the Zel'dovich and galaxy fields beyond the effect of shot noise. (\textit{Bottom}) The ratio of $\rho_{xc}$ to the shot-noise expectation for each sample. Dashed lines represent this value for the real-space power spectra. The real space measurements exhibit less decorrelation beyond $k\sim0.2\ihmpc$, indicating that additional decorrelation is present in redshift space due to, e.g., the finger of god effect.}\label{fig:gal_comp}
\end{figure}

Figure \ref{fig:gal_comp} investigates the extent to which these samples correlate with a Zel'dovich control variate that includes linear and quadratic bias operators. The top panel shows $\rho_{xc}$ for these samples, compared to dashed lines representing each respective samples' shot-noise limit. Here, we see that the samples including satellite galaxies decorrelate from our control variate more rapidly than the central-galaxy-only sample, suggesting that non-linearities in the form of non-linear bias and small scale velocity contributions do lead to additional decorrelation, as expected. Somewhat surprisingly, the amount of additional decorrelation is not particularly sensitive to what multipole we are looking at, although some slight additional decorrelation is apparent for $\ell=2$ with respect to $\ell=0$. The fact that the behavior is so similar between the multipoles can largely be attributed to the fact that $P_{0}(k)$ is the dominant contribution to our disconnected covariances. We have also examined the convergence of our covariance matrices with respect to $\ell_{\rm max}$ in App.~\ref{sec:beta}, demonstrating that our findings are not highly sensitive to our choice of $\ell_{\rm max}=4$.

The bottom panels then show the ratio of $\rho_{xc}$ to the shot-noise limit for each sample, where the dashed lines in the bottom left panel show the equivalent quantity in real space. At $k<0.2 \ihmpc$, the solid and dashed lines are nearly identical, although at higher $k$ the real-space power spectra remain more correlated with our control variate than their redshift-space counterparts. Thus, it is apparent that non-linear velocities do degrade the performance of the ZCV method, but only slightly.

\subsection{Comparison to a linear theory control variate}\label{sec:linear_theory}
A common strategy for mitigating sample variance in simulations that has been used for many years is to divide the measurement of interest by the power spectrum of the linear density field used to initialize the simulation, i.e.
\begin{align*}
    Y = \frac{X}{C}\mu_c
\end{align*}
where $X$ is the simulation measurement in question, $C$ is the linear matter power spectrum measured from the initial conditions of the simulation, and $\mu_c$ is the noiseless linear power spectrum. This estimator, known as the ``ratio control variate'' estimator \citep{mcbook}, is biased because $\langle X/C\rangle\ne\langle X\rangle / \langle C \rangle$.

A similar, but unbiased estimator using the same measured quantities is to use the measured linear power spectrum as a replacement for the ZA in the control variate estimator that we have been using for the rest of this work \citep{Pontzen_2016,mcbook}. It is interesting to consider how well this linear theory control variate performs compared to ZCV.

\begin{figure}
\centering
      \includegraphics[width=\columnwidth]{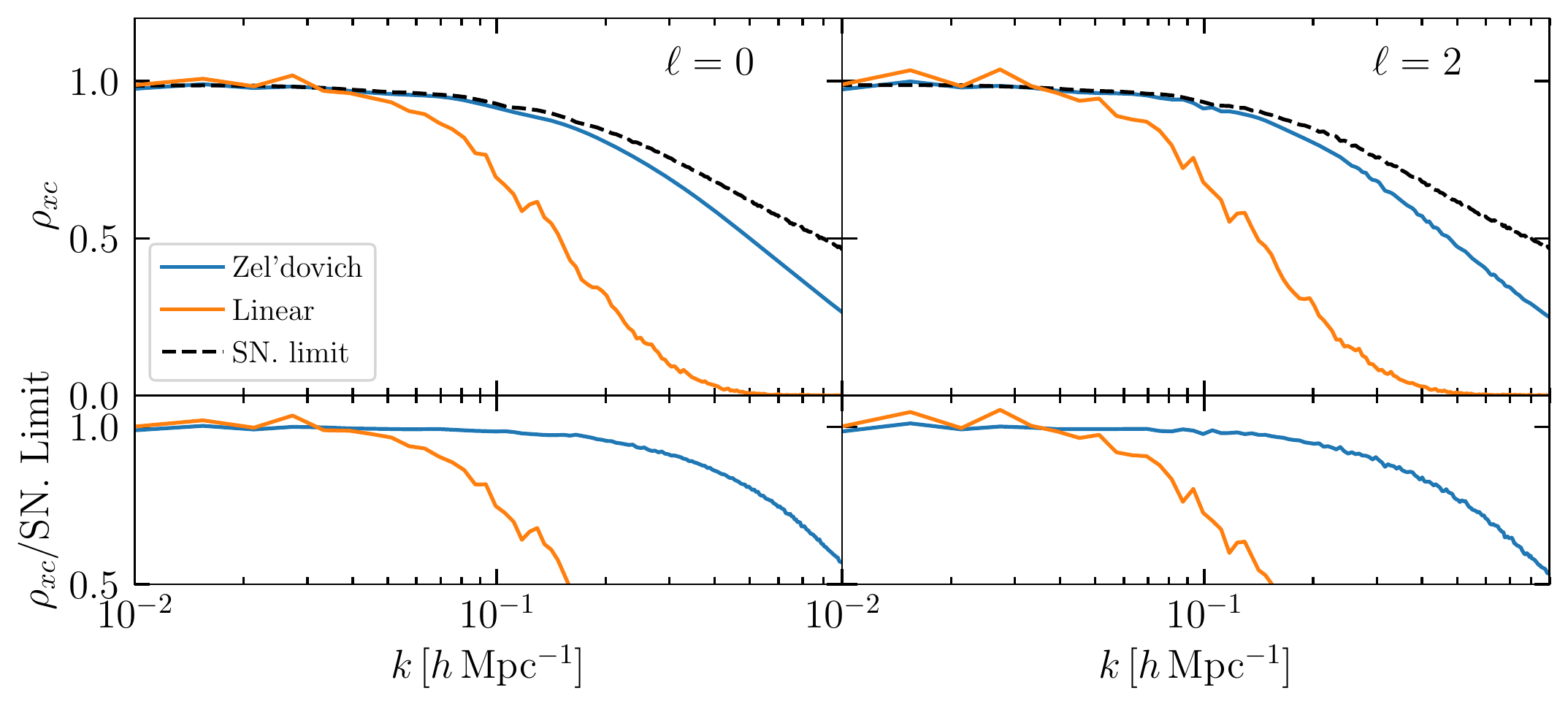}
      \caption{(\textit{Top}) Here we compare the performance of the ZCV method to an alternative linear control variate using the LRG HOD as a test case. We see that the linear control variate performs well out to $k\sim0.05\ihmpc$, but rapidly degrades in performance on smaller scales with respect to ZCV. Furthermore, $\rho_{xc}$ is noisier for the linear control variate, exacerbating the bias discussed in app.\ \ref{sec:beta}. (\textit{Bottom}) A direct comparison of ZCV and linear control variates to the shot-noise limit. Again we see the linear control variate performs significantly worse. }\label{fig:lcv_comp}
\end{figure}

In fig.~\ref{fig:lcv_comp}, we make just such a comparison using our fiducial LRG HOD. We include the Kaiser factor \cite{Kaiser1987}, $(1+b_{\delta_0} + f\mu^2)^2$, in our linear theory control variate in order to optimize its performance. For ZCV, we use a combination of linear and quadratic bias. We see that for scales with $k<0.05\ihmpc$, linear theory performs as well as ZCV, but for $k$ larger than this the performance degrades significantly. Thus, unless one only cares about reducing variance on these very large scales, then ZCV performs significantly better and should be preferred given the small difference in computational cost between the two.

\subsection{Variance reduction forecasts}\label{sec:forecasts}
In the previous sections we demonstrated that we can reduce the variance of biased tracer power spectra in real and redshift space to their shot-noise limit to $k\sim0.2\ihmpc$. Using this fact, we now proceed to forecast the variance reduction that we can expect for a number of galaxy samples of interest for ongoing and upcoming galaxy surveys such as DESI \cite{Aghamousa:2016zmz}, and planned future spectroscopic surveys \cite{Schlegel2022a,Schlegel2022b}. We focus on forecasting two quantities. The first is  $1-\rho_{xc}^2$, which is the effective decrease in variance that is delivered by the ZCV method. We make our forecasts for real- and redshift-space galaxy auto-power spectra as well as for galaxy-matter cross-power spectra. With these, and matter power spectra shown in Fig.~\ref{fig:pmm_example}, we have all the statistics necessary for two-point function analyses of galaxy surveys and their cross-correlations with lensing. We also show $\sigma[P(k)]/P(k)$ assuming a $(2 h^{-1}\rm Gpc)^3$ volume used in this work in order to demonstrate that for the vast majority of galaxy samples, one never needs to simulate a larger volume than this in order to make statistical simulation error subdominant to the systematic error floor associated with current $N$-body codes.

In order to perform these forecasts, we assume that we can achieve the shot-noise limit for $\rho_{xc}$, and thus can predict it given a model for the power spectrum in question, its covariance, and shot-noise level. For simplicity, we assume that only linear bias contributes to the signal, since for $k<0.2\ihmpc$ higher order bias operators contribute at the tens of percent level for reasonable bias values. Furthermore, the covariances are approximated using their disconnected parts, and the shot noise is assumed to be given by the Poisson expectation. We forecast $1-\rho_{xc}^2$ and $\sigma[P(k)]/P(k)$ for five samples: DESI Bright Galaxy Survey (BGS), LRGs, emission line galaxies (ELGs) and quasars (QSOs), as well as for Lyman-alpha emitters (LAEs) which may form the workhorse sample for future spectroscopic surveys \cite{megamapper}. For BGS, we assume $b_{\delta_0}=0$, i.e.\ an Eulerian bias of 1, a number density of $\bar{n}=0.01\, h^{3}\rm Mpc^{-3}$ and an effective redshift of $z_{\rm eff}=0.3$, comparable to the $M_r<-19.5$ sample from SDSS \cite{Zehavi2011}, but at higher redshift. For DESI LRGs \cite{Zhou2022}, we use $b_{\delta_0}=1.2$ and $z_{\rm eff}=0.56$, consistent with the HOD model used in this paper, and $\bar{n}=3\times10^{-4}\, h^{3}\rm Mpc^{-3}$. For ELGs, we take $b_{\delta}=0.7$, $z_{\rm eff} = 1.25$, and $\bar{n}=3.8\times10^{-4}\, h^{3}\rm Mpc^{-3}$ based on \cite{Aghamousa:2016zmz,Raichoor2022}. For QSOs, we assume $b_{\delta}=2$, $z_{\rm eff} = 2$, and $\bar{n}=2.6\times10^{-5}\, h^{3}\rm Mpc^{-3}$ based on \cite{Chaussidon2022}. Finally, for LAEs we set $b_{\delta}=1.5$, $z_{\rm eff} = 2.5$, and $\bar{n}=4\times10^{-4}h^{3}\rm Mpc^{-3}$ \cite{Khostovan2019}. 

\begin{figure}
\centering
      \includegraphics[width=\columnwidth]{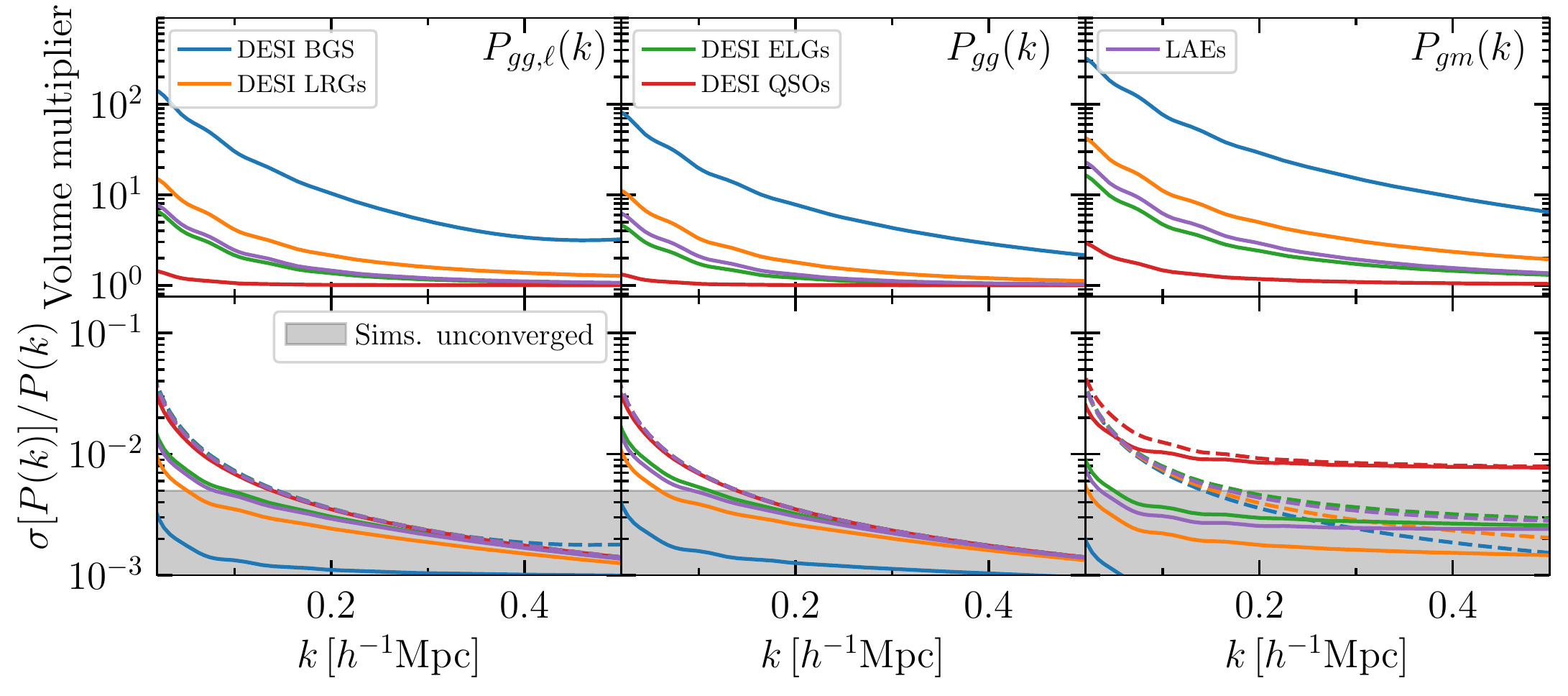}
      \caption{ (\textit{Top}) The factor by which the variance of $\hat{P}_{gg,\ell}(k)$ (left), $\hat{P}_{gg}(k)$ (middle), and $\hat{P}_{gm}(k)$ (right) would be reduced using ZCV for a variety of galaxy samples, as listed in the legend. We have assumed that we can reduce the variance with ZCV up to the shot-noise limit. This assumption should hold to $k\sim0.2\ihmpc$, as seen in fig. \ref{fig:gal_comp}. The number densities, redshifts and linear bias values assumed for these samples are described in the text. (\textit{Bottom}) The fractional error on each statistic for the same galaxy samples as in the top, with (solid) and without (dashed) using ZCV, assuming a volume of $(2 h^{-1} \rm Gpc)^3$. The gray shaded region roughly represents the level at which we expect systematics in simulations to become the dominant error. Thus, for most applications a volume of $(2 h^{-1} \rm Gpc)^3$ is sufficient to remove statistical error from simulated measurements of two-point clustering for galaxy surveys.}

\label{fig:summary_fig}
\end{figure}

Figure \ref{fig:summary_fig} shows the results of these forecasts. The top panel shows $1/(1-\rho_{xc}^2)$, which is the effective increase in volume as a function of scale that the ZCV method provides. We see marked improvements in effective volume for all the samples considered, other than the DESI QSOs which are almost entirely shot-noise dominated at all relevant scales. Unsurprisingly, the sample that performs best is DESI BGS, with its extremely high number density. $P_{gm}(k)$ also exhibits a greater variance reduction for all samples, as the shot-noise contribution to its covariance is reduced with respect to galaxy density auto-correlations. Above $k\sim0.2\ihmpc$, the approximation that ZCV achieves the shot-noise limit on $\rho_{xc}$ may break depending on the satellite fraction and non-linear biasing behavior of the galaxy samples under consideration, and so our forecasts become less trustworthy above these scales.

The solid lines in the bottom panel show the fractional errors that we expect for each sample when applying ZCV, to be compared to the raw fractional errors shown by the dashed lines. The grayed out region roughly represents the level at which we expect systematic errors in our simulations to dominate over statistical errors \citep{Schneider_2016,Garrison2021b,grove2022}. We have assumed a simulation volume of $(2 h^{-1} \rm Gpc)^3$ and a binning of $\Delta k=0.01$ to compute the errors. We see that with this volume we achieve the systematic error floor by $k\sim0.07\ihmpc$ for all the samples considered other than QSOs, thus suggesting that for the majority of galaxy samples that will be observed in the coming years, it will not be important to simulate more than a volume of $(2 h^{-1} \rm Gpc)^3$ in order to remove statistical error from simulated two-point galaxy measurements on scales with $k>0.1\ihmpc$. At larger scales, perturbation theory is known to work exquisitely well \cite{Nishimichi:2020tvu,chen2020redshiftspace}.

\section{Conclusions}
\label{sec:conclusions}

In this work, we have extended the Zel'dovich control variate (ZCV) technique to redshift space. In \cite{Kokron22}, we demonstrated the utility of the Zel'dovich approximation (ZA) as a control variate for real-space measurements. The great benefit of ZCV in real space comes from our ability to inexpensively produce realizations of the ZA, and to analytically predict the mean of those realizations, obviating the two main costs usually associated with the control variate technique. In this work, we showed that we can do so equally well for multipoles of the redshift-space power spectrum of biased tracers, achieving the theoretical limit for variance reduction imposed by the shot noise of those tracers.

In section~\ref{sec:controlvariates}, we gave an overview of the ZCV technique in its simplest form, providing a recipe for applying it to the redshift-space matter power spectrum, which we then expanded to include a treatment for biased tracers in the following sections. In section~\ref{sec:za_rsd}, we described how we make redshift-space power spectrum predictions in the ZA, both on a grid for a particular realization of a linear density field, and for noiseless analytical calculations of the ensemble mean.  Section~\ref{sec:sims} introduced the simulations and HOD that we used to illustrate the ZCV technique. In section~\ref{sec:results}, we proceeded to demonstrate the extent to which ZCV can reduce the variance of biased tracers by measuring the cross-correlation coefficient between our ZA control variates and power spectra of halos and galaxies measured from an $N$-body simulation. This cross-correlation coefficient, $\rho_{xc}$, is the key statistic that determines the performance of the ZCV method, as the amount of variance reduction achieved by ZCV is given by $1-\rho_{xc}^2$.

In section~\ref{sec:biased_tracers} we showed that even with the simplest unbiased ZA control variate we are able to achieve a significant cross-correlation with redshift-space halo power spectra. Adding linear and quadratic bias operators to our ZA control variate significantly improves on this performance for more biased samples, while including tidal bias offers little to no additional gain. Including satellite galaxies somewhat degrades the cross-correlation at $k>0.2\ihmpc$, depending on the satellite fraction of the galaxy sample in question. Nevertheless, when using a Zel'dovich control variate that incorporates linear and quadratic bias, we are able to achieve the theoretical maximum cross-correlation in the presence of shot noise out to $k\sim0.2\ihmpc$ for all the tracers considered in this work. Here we have focused on $z=0.55$ as a fiducial redshift for our tests, but we expect that this performance will improve at higher redshift. At lower redshifts it may degrade, but the volumes probed at redshifts $z<0.5$ are small enough to be simulated at relatively low cost. In section~\ref{sec:linear_theory}, we compared ZCV to a linear theory control variate, showing that ZCV performs significantly better for $k>0.05\ihmpc$

In section \ref{sec:forecasts} we proceeded to forecast the factor by which the variance of clustering measurements made using a number of galaxy samples of relevance to upcoming cosmological measurements can be reduced using ZCV. Figure \ref{fig:summary_fig} shows this variance reduction or equivalently, the effective increase in simulation volume. It also shows the expected fractional error on measurements made from a hypothetical $(2 h^{-1} \rm Gpc)^3$ simulation. When using ZCV, it is clear that running simulations with volumes larger than $(2 h^{-1} \rm Gpc)^3$ is unnecessary for most applications involving two-point functions of biased tracers, and their cross-correlation with matter in either real or redshift space. This has broad implications for a variety of simulation use cases. For example, ZCV will drastically reduce the cost of producing mock galaxy catalogs that have the sufficient statistical precision to test theoretical models at the accuracy required by upcoming surveys. Furthermore, ZCV will allow those wishing to run suites of simulations for the purposes of emulation to run more simulations at higher resolution than would otherwise be possible, thus making the task of interpolating between simulated cosmologies significantly simpler. 

There are still a number of applications that we imagine to be similarly suitable to the ZCV method. Nearest to the current application, one can imagine reducing the variance of reconstructed power spectra or correlation functions \cite{Eisenstein2007,Padmanabhan2009}. It is possible that an optimal variance reduction of such measurements may even be achieved with just linear theory, given that much of the displacement responsible for the decorrelation with the linear initial conditions is removed by the reconstruction algorithm.

Another potential application is to use ZCV to reduce the variance of measurements made in lightcone simulations \cite{Fosalba2015,DeRose2019} in order to produce variance reduced measurements with realistic observational systematics. This will likely require the construction of a pipeline to produce ZA lightcones to pair with their $N$-body counterparts, but otherwise the same ZCV procedure presented here should be appropriate.

Given the limitations on ZCV imposed by shot noise that we have observed here, it is clear that applications to samples with low shot noise are ideal. Measurements of the power spectra of neutral hydrogen, Lyman-alpha flux, or other low shot-noise fields may be particularly suitable for the ZCV method. To what extent these measurements remain cross-correlated with ZA control variates must still be checked.

Finally, higher order correlation functions are a particularly appealing target, both because they have become a common statistic measured in surveys and because of their utility in predicting covariance matrices beyond the disconnected approximation. Going beyond the ZA may be necessary for such applications, and analytic methods for predicting these higher-order functions are numerically complicated to evaluate in LPT. Nevertheless, mean predictions for these statistics in LPT can still be computed via brute force Monte-Carlo, potentially providing a path forward. We leave these additional avenues to future work.

\acknowledgments
The authors thank Pat McDonald for useful conversations during the preparation of this work.
J.D.~is supported by the Lawrence Berkeley National Laboratory Chamberlain Fellowship. S.C.~is supported by the Bezos Membership at the Institute for Advanced Study.
N.K.~is supported by the Gerald J. Lieberman Fellowship.
M.W.~is supported by the DOE.
This research has made use of NASA's Astrophysics Data System and the arXiv preprint server.
This research is supported by the Director, Office of Science, Office of High Energy Physics of the U.S. Department of Energy under Contract No. DE-AC02-05CH11231, and by the National Energy Research Scientific Computing Center, a DOE Office of Science User Facility under the same contract. Calculations and figures in this work have been made using the SciPy Stack \cite{2020NumPy-Array,2020SciPy-NMeth,4160265}. Power spectrum measurements were made with \href{https://github.com/cosmodesi/pypower}{\texttt{pypower}}.

\appendix

\section{Exact redshift-space power spectrum in the Zel'dovich approximation}
\label{app:zenbu}

In ref.~\cite{Kokron22}, the present authors derived exact expressions for the real-space power spectrum within the ZA for tracers up to quadratic order in the bias. Using the cumulant theorem for Gaussian variables, the power spectrum can be put into the form
\begin{equation}
    P(\bk) = \sum_{\mo_{\alpha}, \mo_{\beta}} b_{\mo_{\alpha}} b_{\mo_{\beta}} \int d^3 \bq\ e^{i\bk \cdot \bq - \frac{1}{2} k_i k_j A_{ij}(\bq)}\ F_{\mo_{\alpha}\mo_{\beta}}(\bk,\bq)
    \label{eqn:real_space_general}
\end{equation}
where the sum is over the operators $m$, $\delta_0$, $\delta_0^2$ and $s_0^2$, and the correlator in the exponent $A_{ij} = \langle \Delta_i \Delta_j \rangle$ is the second moment of the pairwise displacement $\Delta = \Psi(\bq) - \Psi(\textbf{0})$. For example, the $(\delta_0^2, s_0^2)$ contribution is given by
\begin{equation}
    F_{\delta_0^2 s_0^2}(\bk,\bq) = 2 E_{ab} E_{ab} - 4 k_i k_j U_i E_{ab} B_{jab} + k_i k_j k_k k_l U_i U_j B_{kab} B_{lab}
\end{equation}
where we are using $i,j,k,l,...$ to denote indices on \textit{displacements} $\Psi_i$ and $a,b,c,d,...$ to denote indices on shears $s_{ab}$ in correlators like $U_i$ and $B_{iab}$. These correlators are functions of the Lagrangian pairwise separation $\bq$ by construction, e.g. $U_i = \langle \Psi_i(\bq) \delta_0(\textbf{0}) \rangle$ is the correlation function between the Lagrangian displacement and the initial density. A comprehensive list of $F_{\mo_{\alpha}\mo_{\beta}}$ are given in Equation A.7 of ref.~\cite{Kokron22}.

The functions $F_{\mo_{\alpha}\mo_{\beta}}$, thus defined, can be decomposed into scalar products of $\hat{k}$ and $\hat{q}$; that is, defining $\mu_{\bq} = \hat{k} \cdot \hat{q}$, we can write
\begin{equation}
    F_{\mo_{\alpha}\mo_{\beta}} = F_{\mo_{\alpha}\mo_{\beta}}^{(0)}(k,q) + F_{\mo_{\alpha}\mo_{\beta}}^{(1)}(k,q)  \mu_{\bq} + F_{\mo_{\alpha}\mo_{\beta}}^{(2)}(k,q) \mu_{\bq}^2 + ...
\end{equation}
Importantly, with the $\mu_{\bq}$ dependence factored out, the coefficients $F_{\mo_{\alpha}\mo_{\beta}}^{(n)}$ are scalar functions of the vector magnitudes $k, q$ only. Similarly, we can write by symmetry $A_{ij}(\bq) = X(q) \delta_{ij} + Y(q) \hat{q}_i \hat{q}_j$. The angular dependence of Equation~\ref{eqn:real_space_general} can then be explicitly separated out as
\begin{equation}
    P(\bk) = \sum_{\mo_{\alpha}, \mo_{\beta}} b_{\mo_{\alpha}} b_{\mo_{\beta}} \sum_n \int dq\ q^2\ K_n(k,q)\ e^{-\frac{1}{2}k^2 X(q)} \ F_{\mo_{\alpha}\mo_{\beta}}^{(n)}(k,q).
    \label{eqn:real_space_general_asep}
\end{equation}
The angular kernel is given by
\begin{equation}
    K_n(q) = \int d\phi\ d\mu_{\bq}\ e^{ikq\mu_{\bq} - \frac{1}{2} k^2 Y(q) \mu_{\bq}^2}\ \mu_{\bq}^n
\end{equation}
independent of $\mo_{\alpha}, \mo_{\beta}$ and can be expressed as a series of spherical bessel functions \cite{Vlah15_loop,Vlah_2016}, and the spherical coordinates are set up such that $\hat{k}$ is at the zenith.\footnote{See Equation F.1 of ref.~\cite{Chen_2020} for an exact expression.}.

As described in Section~\ref{sec:za_rsd}, the only difference incurred by moving to redshift space is that all the displacements must now be multiplied by the matrix $\textbf{R}$ or, equivalently, since all displacements are dotted with wavevectors, all wavevectors other than the one in the Fourier transform ($\bk \cdot \bq$) must be transformed into $K_i = R_{ij} k_j$, such that now \cite{Vlah19,Chen19}
\begin{equation}
    P_s(\bk) = \sum_{\mo_{\alpha}, \mo_{\beta}} b_{\mo_{\alpha}} b_{\mo_{\beta}} \int d^3 \bq\ e^{i\bk \cdot \bq - \frac{1}{2} K_i K_j A_{ij}(\bq)}\ F_{\mo_{\alpha}\mo_{\beta}}(\textbf{K},\bq).
\end{equation}
In this case it is sensible to choose a new system of spherical coordinates with $\hat{K}$ at the zenith; then, the angular dependence can be captured simply by replacing the real-space kernel in Equation~\ref{eqn:real_space_general_asep} with
\begin{equation}
    K_{n,s}(q) = \int d\phi\ d\mu_{\bq}\ e^{i\bk \cdot \bq - \frac{1}{2} K^2 Y(q) \mu_{\bq}^2}\ \mu_{\bq}^n.
\end{equation}
The evaluation of this kernel was performed explicitly in the above references, and the reader is directed to them for further details.

\section{Accounting for incomplete $\mu$ sampling}\label{sec:window}
Measurements of power spectrum multipoles in a periodic box with finite volume incur errors due to mixing between multipoles. This mixing is caused by the sparse sampling in $\mu$ imposed by the discrete nature of the wave modes sampled by the mesh used to measure the simulated power spectra. These artifacts are most significant for large scales and multipoles with $\ell>0$, where the Legendre polynomials become highly oscillatory and so fine $\mu$ sampling becomes more important. This is analogous to the mixing between multipoles (and wavenumbers) caused by incomplete sky coverage in observational data \cite{Beutler2021}, often referred to as the survey window function effect. 

To see how to correct for this effect, we can write our power spectra as 
\begin{align}
    \hat{P}_{\ell}(k_i) &= N_{i}^{-1}\sum_{\bk}B_i(k)\mathcal{L}_{\ell}(\mu)|\delta(\bk)|^2 \\
    &= N_{i}^{-1}\sum_{\bk,\ell^{\prime}}B_i(k)\mathcal{L}_{\ell}(\mu)\mathcal{L}_{\ell^{\prime}}(\mu)P_{\ell^{\prime}}(k)
    \label{eq:pell_est}
\end{align}
\noindent
where $k=|\bk|$, and $\mu=\hat{\mathbf{n}}\cdot\hat{\bk}$, with $\hat{\mathbf{n}}$ the line-of-sight unit vector. $B_i(k)$ is a top-hat around $k_i$, and $N_i = \sum_{\bk}B_i(k)$. The response of this estimator to a change in the uncoupled, noiseless theory, $P_{\ell^{\prime}}(k^{\prime})$, is:
\begin{align}
    W^{i}_{\ell,\ell^{\prime}} &= \frac{\partial \langle \hat{P}_{\ell}(k_i)\rangle}{\partial P_{\ell^{\prime}}(k^{\prime})} = N_{i}^{-1}\sum_{\bk} B_i(k) \mathcal{L}_{\ell}(\mu) \mathcal{L}_{\ell^{\prime}}(\mu)\delta^{D}(k-k^{\prime})\, .
    \label{eq:window_ij}
\end{align}
\noindent
With this window matrix, we can correctly account for the simulation window function when making theory predictions. Alternatively, by assuming that $P_{\ell^{\prime}}(k^{\prime})$ is constant over the width of the bin, $k_i$, then we have
\begin{align}
    \bar{W}^{i}_{\ell,\ell^{\prime}} &= \frac{\partial \langle \hat{P}_{\ell}(k_i)\rangle}{\partial P_{\ell^{\prime}}(k_i)} = N_{i}^{-1}\sum_{\bk} B_i(k) \mathcal{L}_{\ell}(\mu) \mathcal{L}_{\ell^{\prime}}(\mu)\, ,
\end{align}
where $P_{\ell^{\prime}}(k_i) = N^{-1}_i\sum_{k^{\prime}} B_i(k^{\prime})P_{\ell^{\prime}}(k^{\prime})$. Then we have
\begin{align}
    P^{*}_{\ell}(k_i) = (\bar{W}^{-1})^{i}_{\ell,\ell^{\prime}} \hat{P}_{\ell^{\prime}}(k_i)\, ,
    \label{eq:window}
\end{align}
where repeated indices are summed over. $\hat{P}^{*}_{\ell}(k_i)$ is now an unbiased estimator of $\hat{P}_{\ell}(k_i)$ to the extent that our piecewise constant approximation holds and we have included a sufficient number of multipoles in eq. \ref{eq:window}. 

Whether one should deconvolve this effect from the simulation measurements, or convolve the theory with the window matrix depends on the application under consideration. If the simulation measurements are binned finely in $k$, then deconvolution is appropriate, because the assumption that the theory is piecewise constant likely holds to a good approximation. When one wishes to bin more coarsely in $k$, then convolving a finely sampled theory prediction with the window is more appropriate. In this work we deconvolve the window from all the multipole measurements made, in order to ensure that the ensemble mean of our Zel'dovich realizations is equal to the noiseless, uncoupled mean predicted by our analytic ZA model.

\section{Estimating $\beta$ and $\rho_{xc}$}
\label{sec:beta}

We have seen that accurate estimates of $\beta$ and $\rho_{xc}$ are of significant importance, $\beta$ for obtaining an optimal control variate estimator, and $\rho_{xc}$ for accurately predicting the variance reduction factor provided by said control variate. We can see from eqs.~\ref{eq:beta} and \ref{eq:rhoxc} that $\beta$ and $\rho_{xc}$ both rely on accurate estimates of the variance of the control variate, $\textrm{Var}[C]$, and the covariance between the control variate and the statistic whose variance we wish to reduce, $\textrm{Cov}[X,C]$. $\rho_{xc}$ additionally depends on the variance of the latter, $\textrm{Var}[X]$. 

In this work, we have employed the disconnected approximation to estimate covariances, using power spectra measured from our simulations where required. For example, when measuring $\beta$ for real space power spectra, our estimator reads:

\begin{align}
    \beta &= \frac{P^{tz}(k)^2}{P^{zz}(k)^2}
\end{align}
where $\hat{P}^{tz}(k)$ is the measured cross-power spectrum between the tracer in question and our ZA control variate, and $\hat{P}^{zz}(k)$ is the auto-power spectrum of the ZA control variate. Because the tracer field and Zel'dovich field come from the same initial noise realization, the variance of the two spectra largely cancels out, yielding a low-noise estimate of $\beta$. 

\begin{figure}
\centering
      \includegraphics[width=\columnwidth]{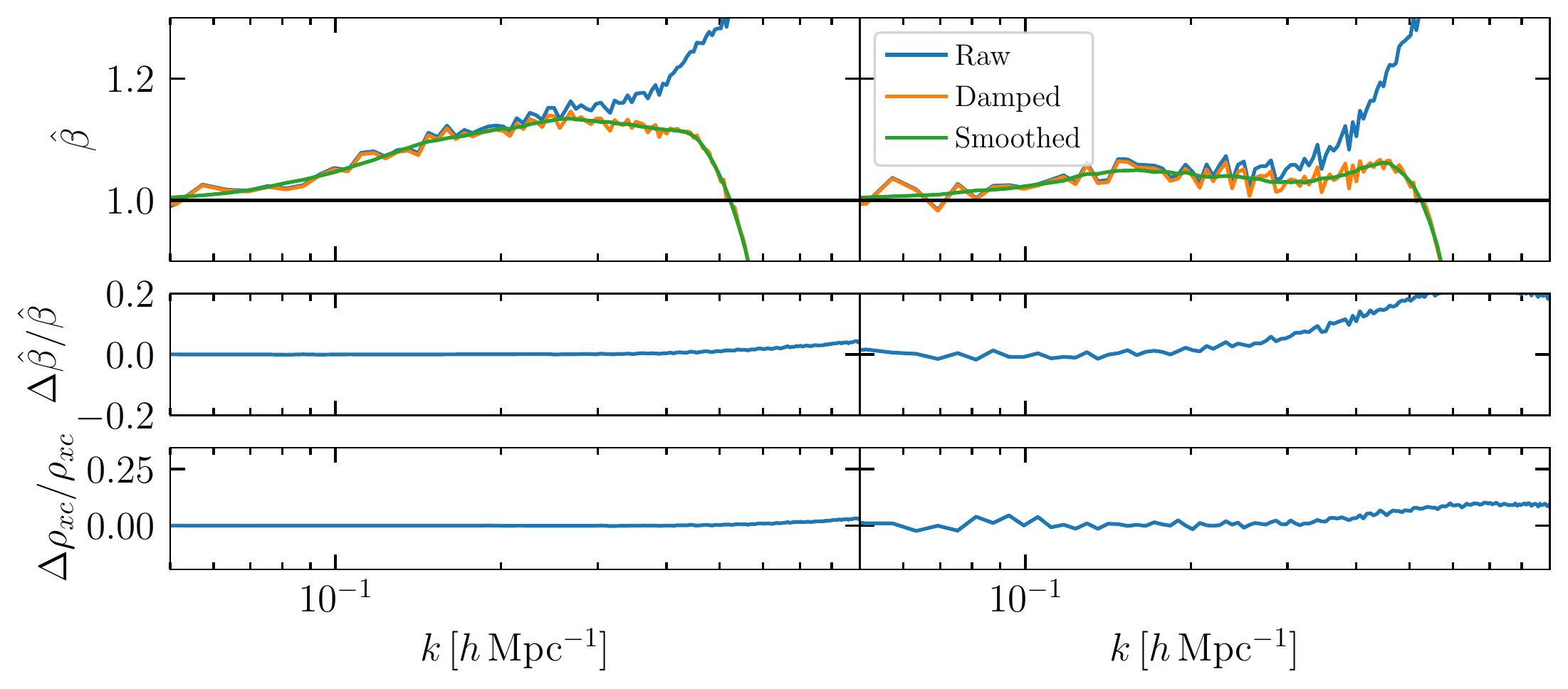}
      \caption{(\textit{Top}) A depiction of our procedure for estimating $\beta$ from our simulations, in the case of the "2x sats." HOD shown in fig.~\ref{fig:gal_comp}. The raw measurements assuming a disconnected covariance approximation are shown in blue. We then apply a damping function (eq.~\ref{eqn:tanhfilter}) to these measurements in order to take $\beta$ to 0 where we expect this disconnected approximation to break down. We additionally smooth $\beta$ with a Savitsky-Golay filter in order to debias our control variate estimator. (\textit{Middle}) The fractional error that is made in $\beta$ by not including the hexadecapole in our covariance estimates. This error is negligible in the regime that we do not damp $\beta$ to zero. Note, that we include the hexadecapole for the rest of this work, and that we expect inclusion of even higher multipoles in our covariance to have even less of an impact. (\textit{Bottom}) Same as the middle panel but for $\rho_{xc}$. Again, the error incurred is negligible. }\label{fig:beta_comp}
\end{figure}

For multipoles of the redshift-space cross-power spectrum between two fields $a$ and $b$, this disconnected covariance reads

\begin{equation}
    \text{Cov}\Big[ P^{aa}_\ell(k), P^{bb}_{\ell'}(k) \Big] = \frac{(2\ell+1)(2\ell'+1)}{V_{\rm obs}} \frac{1}{4\pi k^2 \Delta k} \int d\mu\ \mathcal{L}_\ell(\mu) \mathcal{L}_{\ell'}(\mu)\ P^{ab}(k,\mu)^2
\end{equation}
\noindent 
Avoiding the full generality of Wigner-$3j$ symbols, we can approximate $P^{ab}(k,\mu) = P_0 \mathcal{L}_0 + P_2 \mathcal{L}_2 + P_4 \mathcal{L}_4$ in order to compute the necessary covariances. Thus, our estimates of $\beta$ and $\rho_{xc}$ depend on the maximum $\ell$ that we include in our multipole expansion of $P^{ab}(k,\mu)$. As a cross-check that we are not particularly sensitive to our chosen $\ell_{\rm max}=4$, we have re-estimated $\beta$ and $\rho_{xc}$ setting $\ell_{\rm max}=2$ for the "2x sats." HOD depicted in fig.~\ref{fig:gal_comp}. This should be a worst-case scenario for the tracers considered in this work, as it exhibits the most significant hexadecapole moment. As seen in the bottom two panels of fig.~\ref{fig:beta_comp}, for $k<0.4\ihmpc$ the impact is consistent with zero. For $k>0.4\ihmpc$ we observe $\sim10-20\%$ effects, but these are mitigated by the fact that we apply a damping factor to $\beta$ that we discuss presently. 

The disconnected approximation that we assume for covariances breaks down when non-linear mode coupling becomes significant \cite{Hamilton2006}. This leads to a significant overestimation of $\beta$, compared to the value that we would measure if we had access to the true covariance of the measurements in question. In order to mitigate this and avoid adding extra variance to our simulations at high-$k$ where our ZA surrogate is no longer strongly correlated, we damp $\beta$ to zero by applying a $\tanh$ function:
\begin{align}
\label{eqn:tanhfilter}
    F(k;k_0,\Delta_k) = \frac{1}{2} \left [ 1 - \tanh \left ( \frac{k - k_0}{\Delta_k} \right )   \right ].
\end{align}
\noindent
using $k_0 = 0.618 \ihmpc$ and $\Delta_k = 0.167 \ihmpc$, which were determined by \cite{Kokron22} to bring $\beta$ into agreement with measurement made using an ensemble of simulations. In the future we will redetermine these constants for specific use cases as needed. The upshot of this damping is that we set $\beta=0$ for nearly the same $k$ range as where we observe significant variation when varying $\ell_{\rm max}$.

There is also a subtle bias that enters into the control variate estimator when reducing the variance of measurements from the same simulation that is used to estimate $\beta$. Using real-space auto-power spectra as an example, this can be seen by the fact that the ensemble average  
\begin{align}
    \langle y \rangle &= \langle \hat{P}^{tt}(k) - \hat{\beta} (\hat{P}^{zz}(k) - P^{zz}(k))\rangle \\
      &= \bigg\langle \hat{P}^{tt}(k) - \frac{\hat{P}^{tz}(k)^2}{\hat{P}^{zz}(k)^2} (\hat{P}^{zz}(k) - P^{zz}(k))\bigg\rangle\\
      &= \bigg\langle \hat{P}^{tt}(k) - \left (\frac{\hat{P}^{tz}(k)^2}{\hat{P}^{zz}(k)} - \frac{\hat{P}^{tz}(k)^2}{\hat{P}^{zz}(k)^2}P^{zz}(k)\right)\bigg\rangle \neq \langle \hat{P}^{tt}(k) \rangle 
\end{align}
\noindent 
where hatted variables are measurements from individual realizations, and the ensemble average is taken over many realizations. In order to mitigate this bias, we smooth our estimates of $\beta$ using a third-order Savitsky-Golay filter with a window length of 21. For each $k$ bin, this filter fits a third order polynomial to the 20 adjacent $k$ bins. Thus it is possible to calculate that the effective number of points that each smoothed point receives contributions from is 9.3. In the regime where ZCV yields significant gains, these points are largely uncorrelated. Thus, this process serves to suppress any small correlation between $\beta$ and $\hat{P}^{zz}(k)$ that might prevent $\hat{\beta} (\hat{P}^{zz}(k) - P^{zz}(k))\rangle$ from averaging to zero.\footnote{Let us consider the effect of smoothing within a simple toy model. Consider a set of $N$ random variables $x_i$ with associated control variates $\mu_i$ with zero mean, as well as estimated $\beta_i$, such that we can form the variance-reduced combination $y_i = x_i - \beta_i \mu_i$. Here we imagine $x_i$ is the power spectrum in bin $i$, so for simplicity let us assume $\langle x_i \rangle = \sigma$ and that different bins are uncorrelated; in particular, let us assume that $\beta_i$ and $\mu_i$ can be correlated, as discussed above, such that $\langle \beta_i \mu_j \rangle = c\ \delta_{ij}$. In this case we see that $y_i$ is biased:
\begin{equation}
    \langle y_i \rangle = \sigma - c.
\end{equation}
Now, suppose we instead estimate the correlation by taking the mean across bins, i.e. $\beta = \frac{1}{N} \sum_i \beta_i$. In this case we can see that
\begin{equation}
    y'_i = x_i - \beta \mu_i, \quad \langle y_i \rangle = \sigma - \frac{c}{N},
\end{equation}
i.e. the bias is suppressed by the number of samples. Moreover, we can see that in the case that we are interested in the \textit{overall} amplitude of the $x_i$ (e.g. $\sigma_8$) that
\begin{equation}
    Y = \frac{1}{N} \sum_i y_i, \quad \langle Y \rangle = \sigma - \frac{c}{N}
\end{equation}
similarly has a suppressed bias. It is worth noting however that the additional correlation between $\beta$ and $\mu_i$ will increase the noise of $y'_i$, i.e.
\begin{equation}
    \langle y'_i y'_i\rangle_c \ni \langle \beta \mu_i \rangle^2 = \frac{c^2}{N^2}
\end{equation}
however this too is suppressed by $N$ and, in our case, further by the smallness of the correlation between $\beta$ and $\mu$ as argued in the text.
} Furthermore, since $\hat{P}^{tz,tt}$ are highly correlated, their ratio $\beta$ should be significantly less correlated with $\hat{P}^{zz}$ than each individually, suppressing the bias due to correlations of the control variate with $\beta$. This is why the $\rho_{xc}$'s shown in this paper are far less noisy than the spectra themselves, with noise primarily due to stochasticity on large scales. The process of smoothing and damping $\beta$ described here is depicted in the top panel of fig. \ref{fig:beta_comp}.

Alternatively, if one wishes to avoid these complications and set $\beta$ to some anzatz, e.g. $\beta=1$, as it must be on large scales when including linear and quadratic bias, then the variance of $y$ is
\begin{align*}
    \mathrm{Var}[Y|\beta]
    = \mathrm{Var}[X] + \beta^2 \mathrm{Var}[C] - 2\beta \rho_{xc}\sigma_{x}\sigma_{c}
    = \sigma_c^2 (\beta - \beta^\ast)^2 + \textrm{Var}[Y|\beta^\ast]
\end{align*}
and is thus degraded from the optimal choice of $\beta=\beta^{*}$ by
\begin{align*}
    \frac{\mathrm{Var}[Y|\beta]}{\mathrm{Var}[Y|\beta^{*}]}
    = \frac{1 + \beta^2 \frac{\mathrm{Var}[C]}{\mathrm{Var}[X]}
    - 2\beta\rho_{xc}\frac{\sigma_{c}}{\sigma_{x}}}{1 - \rho_{xc}^2}\,
    = 1 + \frac{\sigma_c^2}{\sigma_x^2}\, \frac{ (\beta-\beta^\ast)^2 }{1-\rho^2} \, .
\end{align*}
If we take the LRG-like HOD as an example, we find $\mathrm{Var}[Y|\beta=1] / \mathrm{Var}[Y|\beta^{*}] \simeq 1.05$ at $k\simeq 0.2\ihmpc$. Thus, one can eliminate the need to estimate $\beta$, apply smoothing, etc., by setting $\beta=1$ without incurring a large penalty in variance, although there is no guarantee of this fact.  This simplifies the pipeline and saves approximately $25\%$ of the computing cost required by ZCV as there is no longer a need to compute cross correlations between the Zel'dovich and tracer fields.

\bibliography{main}
\bibliographystyle{jhep}

\end{document}